\documentclass[amsmath,amssymb,11pt]{article}
\usepackage{jheppub}
\pdfoutput=1
\usepackage{graphicx,epsfig}
\usepackage{float}
\usepackage{subfig}
\usepackage{subfloat}
\usepackage[utf8]{inputenc}
\usepackage{hyperref}
\usepackage{caption}
\usepackage{color}
\usepackage{tensor}
\usepackage{mathtools}
\usepackage{physics}
\usepackage{comment}
\usepackage{pgfplots}
\usepackage{dsfont}
\usepackage[utf8]{inputenc}

\usepackage{hyperref}

\usepackage{afterpage}

\usepackage{graphicx,color,amsmath,amsfonts,enumerate,amsthm,amssymb,mathtools,enumitem,thmtools,hyperref,mathdots,enumitem,centernot,bm,soul,bbm}
\usepackage{braket}
\usepackage{float}
\usepackage{bbold}
\usepackage{comment}
\usepackage{mathrsfs}
\usepackage{tensor}
\usepackage{dsfont}
\usepackage[english]{babel}
\usepackage{tikz,pgfplots}
\usepackage{amsbsy,tensor}
\usepackage{mathrsfs}  
\usepackage{amssymb}
\usepackage[most]{tcolorbox}

\DeclareMathOperator*{\sumint}{%
\mathchoice%
  {\ooalign{$\displaystyle\sum$\cr\hidewidth$\displaystyle\int$\hidewidth\cr}}
  {\ooalign{\raisebox{.14\height}{\scalebox{.7}{$\textstyle\sum$}}\cr\hidewidth$\textstyle\int$\hidewidth\cr}}
  {\ooalign{\raisebox{.2\height}{\scalebox{.6}{$\scriptstyle\sum$}}\cr$\scriptstyle\int$\cr}}
  {\ooalign{\raisebox{.2\height}{\scalebox{.6}{$\scriptstyle\sum$}}\cr$\scriptstyle\int$\cr}}
}

\def\H{ {\cal H} }

\def\L{ {\cal L} }
\def\M{ {\cal M} }
\def\P{ {\cal P} }
\def\R{ {\cal R} }

\newcommand{\Mpl}{M_{\mathrm{Pl}}}

\newcommand*{\affmark}[1][*]{\textsuperscript{#1}}

\newcommand{\beq}{\begin{equation}}

\newcommand{\eeq}{\end{equation}}

\title{Quantum estimation of cosmological parameters}

\author{Michał Piotrak\affmark[1],}
\emailAdd{michal.piotrak.23@ucl.ac.uk}
\author{Thomas Colas\affmark[2],}
\emailAdd{tc683@cam.ac.uk}
\author{Ana Alonso-Serrano\affmark[3,]\affmark[4],}
\emailAdd{ana.alonso.serrano@aei.mpg.de}
\author{Alessio Serafini\affmark[1]}
\emailAdd{a.serafini@ucl.ac.uk}

\affiliation{\affmark[1]Department of Physics and Astronomy, University College London,\\
Gower Street, London, WC1E 6BT, United Kingdom}

\affiliation{\affmark[2] Department of Applied Mathematics and Theoretical Physics, University of Cambridge, Wilberforce Road, Cambridge, CB3 0WA, UK}

\affiliation{\affmark[3,]Institut für Physik, Humboldt-Universität zu Berlin, Zum Großen Windkanal 6, 12489 Berlin, Germany}
\affiliation{\affmark[4]Max-Planck-Institut f\"ur Gravitationsphysik (Albert-Einstein-Institut), \\Am M\"{u}hlenberg 1, 14476 Potsdam, Germany}

\abstract{Understanding how well future cosmological experiments can reconstruct the mechanism that generated primordial inhomogeneities is key to assessing the extent to which cosmology can inform fundamental physics. In this work, we apply a quantum metrology tool — the quantum Fisher information — to the squeezed quantum state describing cosmological perturbations at the end of inflation. This quantifies the ultimate precision achievable in parameter estimation, assuming ideal access to early-universe information. By comparing the quantum Fisher information to its classical counterpart — derived from measurements of the curvature perturbation power spectrum alone (homodyne measurement) — we evaluate how close current observations come to this quantum limit. Focusing on the tensor-to-scalar ratio as a case study, we find that the gap between classical and quantum Fisher information grows exponentially with the number of \textit{e}-folds a mode spends outside the horizon. This suggests the existence of a highly efficient (but presently inaccessible) optimal measurement. Conversely, we show that accessing the decaying mode of inflationary perturbations is a necessary (but not sufficient) condition for exponentially improving the inference of the tensor-to-scalar ratio. 
}


\begin{document}

\maketitle

\section{Introduction}


Parameter inference — the process of estimating the numerical values of a model’s parameters — is a fundamental aspect of scientific inquiry across virtually all disciplines. In cosmology, it plays a central role in shaping our understanding of the Universe, from its composition to its history. Observational data, such as the light captured by telescopes, are translated into quantitative statements about, for example, the abundance of dark matter in the cosmos. This naturally leads to two key questions: how accurately can we infer a given parameter, and what are the optimal methods to do so?


Our current understanding of cosmology traces the origin of structure formation back to quantum fluctuations of the primordial vacuum, seeded during an early phase of accelerated expansion known as inflation \cite{Starobinsky:1979ty,Guth:1980zm,Starobinsky:1980te,Sato:1980yn,Linde:1981mu,Mukhanov:1981xt,Guth:1982ec,Albrecht:1982wi,Starobinsky:1982ee,Hawking:1982cz,Linde:1983gd,Bardeen:1983qw,Mukhanov:1988jd}. A key question is whether — and to what extent — we can reconstruct the quantum state that governs the statistics of these primordial fluctuations. Experimentally, our access to information about the early universe remains limited: we probe it primarily through the lowest-order $n$-point functions of scalar perturbations, known as cosmological correlators. To date, measurements of temperature and polarization anisotropies in the Cosmic Microwave Background (CMB) have confirmed the inflationary prediction of a nearly scale-invariant power spectrum for curvature perturbations \cite{Planck:2018jri,ACT:2025fju,ACT:2025tim,SPT-3G:2025bzu}. However, higher-order statistics — such as the non-Gaussianity of scalar perturbations, the 
$n$-point functions of primordial gravitational waves, and correlators involving the conjugate momentum of the curvature perturbation — remain, at best, weakly constrained, and at worst entirely unmeasured. This limited observational access significantly restricts our ability to infer the quantum state associated with inflation.

In this work, focusing on Gaussian statistics, we aim to characterize the loss of constraining power that results from observing only the power spectrum of the curvature perturbation. 
Consider a single scalar degree of freedom, denoted by the operator $\hat{\mathcal{R}}$, along with its canonical conjugate momentum $\hat{\mathcal{P}}$. The full specification of a Gaussian quantum state requires knowledge of the second moments — that is, the entries of the covariance matrix \cite{serafini2024}
\begin{align}
\sigma_{\mathcal{R}} \equiv \begin{pmatrix}
2\langle \hat{\mathcal{R}}^2 \rangle & \langle \{ \hat{\mathcal{R}}, \hat{\mathcal{P}} \} \rangle \\
\langle \{ \hat{\mathcal{R}}, \hat{\mathcal{P}} \} \rangle & 2\langle \hat{\mathcal{P}}^2 \rangle
\end{pmatrix},
\end{align}
where the expectation values are taken with respect to the quantum state of inflationary perturbations, and $\{A, B\} = AB + BA$ denotes the anticommutator. For a general mixed state, three independent correlators are needed to specify the covariance matrix; in the pure state limit, this reduces to two.
However, as we will derive below, the inflationary dynamics constrain us to observe only the field variance $\langle \hat{\mathcal{R}}^2 \rangle$, which corresponds to the primordial power spectrum. The other entries — $\langle \{ \hat{\mathcal{R}}, \hat{\mathcal{P}} \} \rangle$ and $\langle \hat{\mathcal{P}}^2 \rangle$ — require access to the so-called decaying mode, associated with the velocity field $\dot{\mathcal{R}}$, which becomes exponentially suppressed with the number of \textit{\textit{e}-folds} $N_{\mathrm{ef}}$ that a given mode spends outside the horizon. For CMB-relevant scales, this suppression typically occurs over $N_{\mathrm{ef}} \sim 50$.
This raises a natural question: \textit{how much more information could be extracted if we had access to statistics involving the conjugate momentum?}


Fisher forecasts are a standard tool in the analysis pipelines of upcoming cosmological surveys~\cite{Font-Ribera:2013rwa,Euclid:2019clj,Sohn:2019rlq}, used to anticipate how well future experiments will be able to constrain theoretical parameters. In this work, we adopt a related technique from quantum metrology — the \textit{Quantum Fisher Information} (QFI) — to estimate the ultimate precision with which microphysical parameters of inflationary models could, in principle, be inferred. We adapt this formalism to general squeezed states, which notably include the quantum state in which cosmological inhomogeneities are placed during inflation. 
As a concrete application, we use this framework to estimate the precision on the first slow-roll parameter, which directly controls an observable quantity: the tensor-to-scalar ratio of primordial perturbations. Finally, by comparing the QFI with the classical Fisher information obtained from observations of the curvature perturbation power spectrum alone — that is, from measurements of $\langle \hat{\mathcal{R}}^2 \rangle$ — we assess whether the measurement imposed by nature is close to optimal, or if it leaves significant untapped information.


This work emerges from the growing interest in the interplay between quantum information and high-energy physics~\cite{Calabrese:2004eu, Rosenhaus:2014woa, Peschanski:2016hgk, Howl2021, Cheung:2023hkq, Aoude:2024xpx,  Agullo:2023fnp, Agullo:2024har, Fedida23, Fedida25, Asenov:2025ueu, ignoti, Subba:2024mnl, Shivashankara:2024oue, Tajima:2024tzu, Lin:2024roh, Hentschinski:2024gaa, Kaku:2024lgs, Boutivas:2025ksp, Bernal:2024xhm, Afik:2024uif, Guimaraes:2024lqf, Wu:2024ues, Noumi:2025lbb, Cai:2025kcd, Aoude:2025jzc, Guimaraes:2024mmp, Assi:2025ibi, Han:2024ugl, Afik:2025ejh, Kong:2025ksx, MacIntyre:2025cmr, Sou:2025tyf, Gondret:2025fdc, Salcedo:2024smn, Colas:2024lse, Salcedo:2024nex, Salcedo:2025ezu, Colas:2025app, Burgess:2024heo}. Within primordial cosmology, it contributes to ongoing efforts to characterize the quantum state of inflationary perturbations using tools such as quantum squeezing~\cite{Grishchuk:1990bj,Grishchuk:1992tw,Albrecht:1992kf, Polarski:1995jg, Lesgourgues:1996jc, Kiefer:1998qe, Grain:2019vnq,Hsiang:2021vgx, Martin:2022kph}, quantum discord~\cite{Martin:2015qta,Kanno:2016gas,Martin:2021znx, Martin:2022kph, Micheli:2023qnc}, Bell tests \cite{Campo:2005sv, Maldacena:2015bha, Choudhury:2016cso, Martin:2016tbd, Martin:2017zxs, Ando:2020kdz, Espinosa-Portales:2022yok, Tejerina-Perez:2024opu, Sou:2024tjv} and Clauser–Horne–Shimony–Holt inequalities~\cite{Martin:2016nrr, Ando:2020kdz}, entropy and other quantum information measures~\cite{Brahma:2023hki, Wu:2022xwy, Wu:2022lmc, Wu:2023spa, Colas:2022hlq, Colas:2022kfu, Colas:2023wxa, Colas:2024xjy, Burgess:2024eng, Colas:2024ysu, Katsinis:2024sek, mozaffar2024capacity, Ribes-Metidieri:2024vjn, Ribes-Metidieri:2025nfw}, complexity~\cite{Bhattacharyya:2024duw, Brahma:2024sie, Bhattacharyya:2024rzz, Li:2024iji}, Quantum Speed Limit \cite{Kobayashi:2025pfy} and, now, quantum Fisher information too~\cite{Gomez:2021yhd, Gomez:2020xdb, Gomez:2021hbb, Carrion:2024lap, Chen:2024dcc, Balut:2024aru, Ferro:2025wbw, Frigerio:2025ewb,Banerjee_2018, Iyen:2025htn, Lopez-Pardo:2025eqe}.   

The main assumptions underlying this study are as follows. We restrict ourselves to the Gaussian statistics of scalar perturbations generated during inflation. Our analysis follows the framework of quantum field theory in curved spacetime, assuming a fixed classical background on which quantum fluctuations evolve, thereby neglecting backreaction effects. We consider a quasi-de Sitter background, where deviations from exact de Sitter expansion are controlled by a hierarchy of small slow-roll parameters. All computations are performed at leading order in the slow-roll expansion.


The first two Sections of this manuscript aim to provide sufficient background to the readers of each community to apprehend the main content of this project. Sec.~\ref{sec:QI} reviews symplectic methods in QFT while Sec.~\ref{chap:scalar} presents the current status of primordial cosmology. Readers familiar with these concepts can skip these sections. In Sec.~\ref{sec:QFI}, we introduce techniques from Gaussian parameter estimation and compute the quantum Fisher information for a generic squeezed state. These results are applied to the squeezed states in which are placed cosmological inhomogeneities during the early universe in Sec.~\ref{sec:appinfl}, from which we draw conclusions for primordial cosmology in Sec.~\ref{sec:conclu}. The article is complemented by a series of appendices, in which technical details are deferred.

\paragraph{Notations and units.} Primes denote derivatives with respect to the conformal time while $_{,\vartheta}$ derivatives with respect to a parameter. The main scalar degree of freedom of interest in this work, the curvature perturbation, is denoted $\mathcal{R}$, while $\mathcal{P}$ stands for its conjugate momentum. The tensor perturbations are denoted $\gamma_{ij}$. At last, the symbol $\sigma$ represents covariance matrices. We work in natural units in which $\hbar = c= G_N = 1$ and only use the reduced Planck mass $\Mpl$.

	\section{Symplectic geometry in field theory}\label{sec:QI}
	
	The main mathematical tool used throughout this paper is symplectic geometry, which informs Gaussian quantum information. Indeed, due to the symplectic nature of field theory there is a natural relation between QFT and Gaussian quantum information bridged by symplectic geometry. In this section, we demonstrate how these fields are related. We keep the respective introductions short and for a more detailed and formal explanations we direct the reader to Ref.~\cite{Arnold:1989who} for symplectic geometry and Ref.~\cite{serafini2024} for Gaussian quantum information.
	
	\subsection{Symplectic geometry}

    Throughout this section, we shall restrict our considerations to a non-interacting scalar field theory. Note that they do, however, generalise to other theories. Consider a real scalar field $\Phi$. All possible configurations of that field form a smooth manifold $\mathcal{M}=\{\phi(x):\Sigma_t\to\mathds{R},\, \text{smooth}\}$, where $\Sigma_t\subset \mathcal{N}$ is a Cauchy hypersurface on a $d-$dimensional Lorentzian space-time with coordinates $x$. One may notice that $\mathcal{M}$ is not finite-dimensional. This is not a problem as one may generalise the idea of manifolds to ones with infinite dimension. The tangent space here is an infinite-dimensional vector space, commonly a Fréchet space. While the details lie beyond the scope of this paper, the key point is that the tools of symplectic geometry remain applicable in our setting.

    To each configuration $\phi(x)$ one may associate conjugate canonical momenta $\pi(x)$. Since we have not fixed our Hamiltonian yet, $\pi(x)$ denote all possible directions of evolution of $\phi(x)$ and form an infinite-dimensional vector space, called the cotangent space. A union over all $\phi(x)$ of cotangent spaces gives a cotangent bundle. To put it more formally, we build our covectors in the cotangent bundle from objects like: $\int_{\Sigma_t}\dd^{d} x \pi(x) \mathbf{d}\phi(x)$, where $\mathbf{d}$ denotes a $1-$form on $\mathcal{M}$. Since the cotangent bundle has even dimension (infinite but ``even" as it is a union of cotangent spaces), we can choose a symplectic structure on it, which in coordinates shall be given by \cite{Arnold:1989who}
	\begin{equation}
		\omega=\int_{\Sigma_t}\dd^{d} x \mathbf{d}\pi(x)\wedge\mathbf{d}\phi(x).
	\end{equation}
	The physical meaning of $\omega$ is that it gives a notion of the Poisson bracket between functionals, which is typically given by \cite{Arnold:1989who}
	\begin{equation}\label{eq:Poisson}
		\{f,g\}=\int_{\Sigma_t}\dd^{d} x \left(\frac{\partial f}{\partial \phi(x)}\frac{\partial g}{\partial \pi(x)}-\frac{\partial f}{\partial \pi(x)}\frac{\partial g}{\partial \phi(x)}\right).
	\end{equation}
	In our geometrical setting this is given by
	\begin{equation}
		\{f,g\}=\omega(X_f,X_g),
	\end{equation}
    where $X_h$ denotes a vector field constructed from a functional $h$.
    
	For example, the vector fields associated with the canonical variables are
	\begin{align}
		\begin{split}
			&X_\phi = - \frac{\partial}{\partial \pi(x)},\\
			&X_\pi =  \frac{\partial}{\partial \phi(x)},
		\end{split}
	\end{align}
	and the symplectic form yields the correct algebraic structure through Poisson brackets
	\begin{equation}
		\{\phi(t,x),\phi(t,y)\}=\{\pi(t,x),\pi(t,y)\}=0, \quad \{\pi(t,x),\phi(t,y)\}=(2\pi)^{d}\delta^{d}(x-y).
	\end{equation}
    We see that the purpose of the symplectic form is to introduce an algebraic structure under which the algebra of fields is closed. This is indeed vital for quantization.
	
	Consider now a function $H$, called the Hamiltonian. Any observable $f$ in field theory can be thought of as a curve on the manifold $\mathcal{M}$ whose direction is determined by the Hamiltonian vector field $X_H$. Such curves solve
    \begin{equation}
		\dot{f}=\omega(X_f,X_H).
	\end{equation}
    This can be thought of as a $1-$parameter group of diffeomorphisms\footnote{Differentiable transformation between manifolds which are smooth and have smooth inverses.}, whose generator is the Hamiltonian vector field. Furthermore, the group preserves the symplectic form. We shall call such transformations symplectomorphisms.
	In particular, consider the evolution of $\pi(x),\phi(x)$ 
	\begin{align}
		\label{eq:ham_flow}
		&\dot{\pi}=\omega(X_{\pi},X_H)=-\frac{\partial H}{\partial \phi}=-X_{\pi}(H),\\
		&\dot{\phi}=\omega(X_{\phi},X_H)=\frac{\partial H}{\partial \pi}=-X_{\phi}(H).
	\end{align}
	These are the well-known Hamilton's equations of motion.

    Upon quantisation, the symplectic structure given by $\omega$ gives rise to the algebraic structure of quantum field theory through the usual mapping of Poisson brackets to the canonical commutation relations. Then Hamilton's dynamics simply become Heisenberg's equations of motion. Solving them typically leads to a natural separation into positive and negative frequency modes from which we build the Fock space. This, however, is harder for time-dependent Hamiltonians, i.e. whenever energy is not conserved. In inflationary cosmology there is no time-like Killing field and thus no energy conservation. We must then use a different approach.
    
    Consider again the equations of motion in Eq.~(\ref{eq:ham_flow}). In the case of quadratic Hamiltonians, i.e. functionals where the highest order of field variables is two, the above equations form a linear system of differential equations. They can be written as a matrix differential equation and admit a solution:
    \begin{equation}\label{eq:symplevv2}
        \begin{pmatrix}
            \phi(t)\\
            \pi(t)
        \end{pmatrix}=S(t,t_0)\begin{pmatrix}
            \phi(t_0)\\
            \pi(t_0)
        \end{pmatrix},
    \end{equation}
where the matrix $S(t,t_0)$ belongs to $Sp(2, \mathds{R})$, i.e. the real symplectic group in two dimensions (or its complex representation depending on the exact choice of field variables). We can quantise as usual by mapping the symplectic form onto commutation relations and promoting the field variables to quantum operators. We still need to construct the Fock space, though. This can be done by rotating field operators onto creation/annihilation operators, i.e. expressing $\phi,\pi$ as linear combinations of $a,a^\dagger$. Such rotation is also a symplectic transformation and preserves the commutation relations, as well as the equations of motion. The dynamics for the equations of motion are often referred to as the Bogoliubov transformation. Then the Hilbert space is constructed as usual. We thus arrive at a well-defined quantum field theory with full dynamics encoded in a two-dimensional vector space. Thus, despite the infinite dimension of the Hilbert space, quantum information is possible, as we shall see in the next section.
This exact procedure is presented for a flat space-time real Klein-Gordon field in Appendix~\ref{app:app0}. We also demonstrate it in the cosmological setting in Sec.~\ref{quant}.

\subsection{The symplectic group}

The Bogoliubov transformation (\ref{eq:symplevv2}) we used to solve the Klein-Gordon equation encodes the dynamics of the field and belongs to the symplectic group $Sp(2,\mathds{R})$ (it can, in fact, be turned real through the unitary transformation that moves to quadrature variables). This is a general result: the dynamics of any free theory can always be represented by a symplectic matrix directly corresponding to the symplectomorphism given by the Hamiltonian flow. As a matrix group, real symplectic transformations $S$ in dimension $2n$ (i.e., over $n$ degrees of freedom) are defined by the property:
\begin{equation}
    S\Omega S^T=\Omega,
\end{equation}
where $\Omega=\displaystyle\bigoplus_{i=1}^n \begin{pmatrix}
    0 & 1\\
    -1 & 0
\end{pmatrix}$ is the matrix representation of the anti-symmetric symplectic form. Upon quantization, symplectic tranformations are therefore linear transformations on the field operators that preserve the canonical commutation relations. 

In this paper, we will just deal with single-mode ($n=1$) symplectic transformations, whose most general form can be derived by their singular value decomposition as \cite{serafini2024}:
\begin{equation}\label{svd}
    S = R(\theta) Z(r) R(\varphi) ,
\end{equation}
where $R(\theta)=\left(\begin{array}{cc}\cos\theta&\sin\theta\\-\sin\theta&\cos\theta\end{array}\right)$ and $Z(r)={\rm diag}({\rm e}^{r},{\rm e}^{-r})$. Note that single-mode real symplectic transformations are just real two-times-two matrices with determinant one.

Our evaluations of quantum and classical Fisher information will rely on the mathematical expediency of Gaussian states, a class of states defined as the ground and thermal states of quadratic Hamiltonians (i.e., of Hamiltonians comprised of terms of order two or less in the field operators). Such states admit phase-space descriptions as multivariate Gaussian characteristic or quasi-probability functions \cite{serafini2024}, a property which we will exploit later on to determine exactly their quantum Fisher information.
In closed system such as the one at hand, a single-mode covariance matrix (CM) $\sigma$ that determines the quantum state (along with first moments which are zero in our case) is obtained by applying a symplectic $S$ by congruence on the vacuum's CM (the identity):
\begin{equation}
    \sigma = S S^{\sf T} = R(\theta) Z(2r) R(-\theta) \, ,
\end{equation}
where we applied the singular value decomposition (\ref{svd}).
The expression of the pure Gaussian state $\hat{S}(r)\ket{0}$ with CM $Z(2r)$ (a pure, single-mode squeezed state, obtained by acting with the squeezing transformation $\hat{S}(r)$ on the vacuum state) in the Fock basis will also be useful in the following \cite{serafini2024}
\begin{equation}\label{eq:sqtr}
    \hat{S}(r)\ket{0} = \frac{1}{\sqrt{\cosh(r)}}\sum_{m=0}^{\infty}\left(\frac{\tanh(r)}{2}\right)^m\frac{\sqrt{(2m)!}}{m!}\ket{2m} \; .
\end{equation}

Finally, another ingredient from Gaussian information we will employ is the general-dyne description of Gaussian measurements, encompassing the well-known homodyne case (i.e., the measurements of canonical quadratures, which are applicable in practice in cosmology), the heterodyne case, as well as all intermediate cases. %
Let us first remind the reader that general quantum measurements are described by positive operator valued measurements (POVMs), that is, by sets of positive operators $\Pi_\mu$ such that 
\begin{equation}
    \sumint_{\mu} \Pi_\mu = {\mathds 1} \; ,
\end{equation}
where $\mu$ labels the measurement outcome and we use the notation $\sumint$ to indicate that the sum may be generalised to an integral (whence the `POVM' terminology). For a given quantum state $\varrho$, the probability associated to a measurement outcome $\mu$ is extracted through the so-called Born rule
\begin{equation}
    p(\mu) = {\rm Tr}\left[\varrho \Pi_\mu\right]\, .
\end{equation}
Now, the set of Gaussian, `general-dyne' measurements on a single canonical degree of freedom is described by the POVM
\begin{equation}
    \frac{1}{4\pi} \int_{{\mathbbm R}^2} {\rm d}{\bf r} \hat{D}_{-\bf r} \varrho_G \hat{D}_{\bf r} = {\mathds 1}, 
\end{equation}
where $\varrho_G$ is a Gaussian state with null first moments, while $\hat{D}_{\bf r}$ is the Weyl displacement operator $\hat{D}_{\bf r} = {\rm e}^{i{\bf r}^{\sf T}\Omega \hat{\bf r}}/\sqrt2$ for $\hat{\bf r}=(\hat{\phi},\hat{\pi})^{\sf T}$ (recall that the canonical commutation relations in natural units read $[\hat{\phi},\hat{\pi}]=i$). Since these POVM's describe projections onto general Gaussian states, these measurements preserve the Gaussianity of the state, i.e., in the ideal scenario where the measurement is enacted through a non-demolition scheme, the update corresponding to any outcome $\mu$ will be a Gaussian state.

Such measurements may be described exactly on Gaussian states and result in the following normalized Gaussian probability distributions \cite{serafini2024}:
\begin{equation}
    {\rm d}p({\bf r}) = \frac{{\rm e}^{-\frac12({\bf r}-{\bf \Bar{{r}}})^{\sf T}(\sigma+SS^{\sf T})({\bf r}-{\bf \Bar{{r}}})}}{2\pi \sqrt{{\rm Det}[\sigma+SS^{\sf T}]}}{\rm d}{\bf r} , 
\end{equation}
where ${\bf \Bar{r}}$ is the mean of the measured state, $\sigma$ is its CM, ${\rm d}{\bf r}={\rm d}r_1{\rm d}r_2$ and $S$ is a symplectic operation characterizing the choice of measurement. Homodyne detection is described by the limits $r\rightarrow \infty$, where $r$ is the squeezing parameter of $S$, while $r=0$ corresponds to heterodyning (projection on coherent states). 
Notice that the general-dyne outcomes, denoted with ${\bf r}=(r_1,r_2)^{\sf T}$ above, are usually double-valued, except in the homodyne limit, where only the real outcome with finite (as opposed to infinite) variance survives. In fact, homodyne measurements correspond to the measurement of the hermitian operators $\cos\theta \phi + \sin\theta\pi$, for some phase rotation $\theta$. In what follows, these general-dyne probabilities will allow us to evaluate specific instances of classical Fisher information and contrast them with the optimized quantum Fisher information.


\section{Early universe cosmology}
\label{chap:scalar}

After a brief introduction to the main assumptions underlying the current paradigm in modern cosmology, we review the background evolution in the early universe and the dynamics of the fluctuations therein. Familiar readers can jump straight to the end of Sec.~\ref{subsec:pert} where we introduce the squeezed state describing cosmological inhomogeneities we use through the rest of the article.

Cosmology in the modern sense is a precision discipline combining the analysis of a astonishingly large number of observations (position of the galaxies in the sky, temperature anisotropies of the Cosmic Microwave Background and more recently gravitational waves from binaries and Pulsar Timing Arrays) with theoretical predictions at the crossroad of General Relativity and Particle Physics. Its current paradigm is known as the $\Lambda$CDM model (which stands for cosmological constant and cold dark matter), relying on the following seven assumptions \cite{Planck:2018nkj}: 
\begin{enumerate} 
    \item Physics is the same throughout the observable universe.
    \item General Relativity is an adequate description of gravity.
    \item On large scales, the universe is statistically homogeneous and isotropic.
    \item There are five basic cosmological constituents: dark energy i.e. the cosmological constant which behaves as the energy density of the vacuum, dark matter, atomic matter, photons and neutrinos. 
    \item The curvature of space is very small.
    \item The topology is trivial.
    \item Variations in density of these constituents were set at the early times and are Gaussian, adiabatic and nearly scale invariant, as predicted by an early phase of inflation. 
\end{enumerate}
We work within this framework in the rest of the article and aim at bringing new tools to characterise the early phase of inflation seeding the initial condition for the subsequent evolution of the universe.

\subsection{Background evolution}

The assumption of homogeneity and isotropy yields to consider the Universe described by a Friedmann-Lemaître-Robertson-Walker (FLRW) metric as
\begin{equation}
    \dd s^2=-\dd t^2+a^2(t)\left[\frac{\dd r^2}{1-\kappa r^2}+r^2(\dd \theta^2 +\sin^2\theta \dd \phi^2)\right],
\end{equation}
where $a$ is the scale factor determining the expansion of the Universe by giving the size of the spacelike hypersurfaces at each time $t$, and $\kappa$ is the curvature parameter taking values ${+1, 0, -1}$ for positively curved, flat and negatively curved spacetimes respectively. Sticking to the above assumptions underlying the $\Lambda$CDM model, we will set $\kappa = 0$ for the rest of this work. To characterize the expansion of the Universe, we introduce the Hubble parameter defined as
\begin{equation}
    H\equiv \frac{\dot{a}}{a},
\end{equation}
with units of inverse time. Note that the comoving Hubble radius, $(aH)^{-1}$, is often the relevant scale to compare to a wavenumber of interest $k \sim \lambda^{-1}$. Two regime will play a crucial role below, the \textit{sub-Hubble} regime in which the mode $k \gg aH$ (i.e. $\lambda \ll (aH)^{-1}$) is well within the cosmological horizon, and the \textit{super-Hubble} regime where the mode $k \ll aH$ (i.e. $\lambda \gg (aH)^{-1}$) ``perceives'' the spacetime curvature.

Within the $\Lambda$-CDM model, the early universe is driven by a single scalar field $\phi$, the inflaton, evolving in a flat potential. We will refer to this class of models as \textit{single-field slow-roll inflation}. This phase is designed to permit an early era of accelerating expansion ($\ddot{a}>0$) in order to set up the initial conditions for the subsequent Hot Big Bang. It turns out that having $\ddot{a}>0$ is equivalent to consider the energy budget of the universe being dominated by a perfect fluid with an equation of state $w \equiv p/\rho < -1/3$, neglecting the cosmological constant term which is assumed to be subdominant in the early universe \cite{Dodelson:2003ft}. Let us first review how single-field slow-roll inflation can achieve this task. The Lagrangian density for the scalar field is given by 
\begin{align}
	\mathcal{L}_m = \frac{1}{2} g^{\mu \nu} \partial_\mu \phi \partial_\nu \phi - V(\phi),
\end{align}
from which we deduce the stress-energy tensor
\begin{align}
	T^{(\phi)}_{\mu \nu} = \partial_\mu \phi \partial_\nu \phi - g_{\mu \nu} \left[ g^{\alpha \beta} \partial_\alpha \phi \partial_\beta \phi + V(\phi)\right].
\end{align}
At the background level, for a homogeneous field configuration $\phi(t, x^i) = \bar{\phi}(t)$, one can assimilate the scalar field to a perfect fluid of energy density and pressure
\begin{align}
	\rho &= \frac{1}{2} \dot{\phi}^2 + V(\phi), \\
	p&=\frac{1}{2} \dot{\phi}^2 - V(\phi).
\end{align}
When the potential energy dominates over the kinetic term, $w \simeq -1 < -1/3$ and the universe undergoes a phase of accelerating expansion. This regime in which $\dot{\phi}^2 \ll V(\phi)$ is known as the \textit{slow-roll} regime. It is characterised by a series of moments known as the \textit{Hubble flow parameters} defined as
\begin{align}
	\epsilon_0 &= \frac{H_{\mathrm{ini}}}{H} \quad \quad \text{and} \quad \quad
	\epsilon_{n+1} = \frac{\dd }{\dd N_{\mathrm{ef}}} \ln \epsilon_{n},
\end{align}
where $H_{\mathrm{ini}}$ is some initial value for the Hubble parameter and $N_{\mathrm{ef}} = \ln a$ is the \textit{number of \textit{e}-folds}, which amounts for the number of exponential expansions of the universe. Explicitly, the first two parameters are
\begin{align}\label{eq:srsf}
	\epsilon_1 = - \frac{\dot{H}}{H^2} = \frac{\frac{3\dot{\phi}^2}{2}}{\frac{\dot{\phi}^2}{2} + V(\phi)} \quad \quad \text{and}\quad \quad \epsilon_2 = \frac{\dot{\epsilon}_1}{H \epsilon_1}= 2 \left(\epsilon_1 +  \frac{\ddot{\phi}}{H \dot{\phi}}\right),
\end{align}
where we used the background Einstein equations known as the Friedmann equations, 
\begin{align}
	H^2 = \frac{\frac{\dot{\phi}^2}{2} + V(\phi)}{3M^2_{\mathrm{Pl}}} \quad \quad \text{and} \quad \quad \dot{H} = - \frac{\dot{\phi}^2}{2M^2_{\mathrm{Pl}}},
\end{align}
to relate the Hubble parameter and its derivatives with the kinetic and potential energy of the field. One can easily check that $\dot{\phi}^2 \ll V(\phi)$ imposes $\epsilon_1 \ll 1$, that is $\dot{H}\ll H^2$. Said it differently, the Hubble parameter is enforced to be almost a constant, corresponding to a nearly exponential expansion of the scale factor $a(t)$. This is the reason why inflation is often called a \textit{quasi de-Sitter} expansion, as it only departs from the forever exponentially expanding universe (\textit{de Sitter} spacetime) by a set of slow-roll parameters. Hence, the slow-roll expansion amounts to assume an almost frozen dynamics, for which $|\epsilon_n| \ll 1$, $\forall n \geq 1$. 

In particular, it implies that the scalar field equation of motion given by the Klein-Gordon equation
\begin{align}
	\ddot{\phi} + 3 H \dot{\phi} + V' = 0,
\end{align}
where $V' \equiv \dd V / \dd \phi$, simplifies to its attractor solution (independent of the initial conditions)
\begin{align}\label{eq:attractor}
	\dot{\phi} = -\frac{V'}{3H},
\end{align}
as $|\epsilon_2| \ll 1$ amounts to neglect the acceleration of the field over the other terms of the Klein-Gordon equation. If we finally relate these slow-roll parameters to the shape of the potential using Eq.~\eqref{eq:attractor}, we find 
\begin{align}\label{eq:srpot}
	\epsilon_1 \simeq \frac{M_{\mathrm{Pl}}}{2} \left( \frac{V'}{V}\right)^2\quad \quad \text{and} \quad \quad\epsilon_2 \simeq 2 M_{\mathrm{Pl}} \left[\frac{1}{2} \left( \frac{V'}{V}\right)^2 
	- \frac{V''}{V} \right].
\end{align}
To summarize, having a quasi de Sitter expansion with nearly constant $H$ requires $\epsilon_1 \ll 1$ which implies a flat potential. Reaching the attractor (one-dimensional phase space) trajectory of the form of Eq.~\eqref{eq:attractor} requires $|\epsilon_2| \ll 1$ which requests a small mass for the inflaton $V'' \ll H^2$. Despite being  notoriously known as a difficult task to achieve due to radiative corrections generating an inflaton mass heavier than the Hubble parameter $H$, the smallness of $\epsilon_2$ is a well-tested observational feature. In particular, it directly connects to the nearly scale invariance of the power spectrum which is tested with exquisite precision in the CMB \cite{Planck:2018vyg}. On the contrary, the first slow-roll parameter $\epsilon_1$ is so far only bounded from the lack of observation of primordial gravitational waves. Accessing this parameter experimentally would bring deep insight into the inflationary phase, in particular the energy scale at which it took place. Hence, in the rest of this work, we aim at characterising the precision of inference of this parameter from the state of the perturbations seeded at the end of inflation.

\subsection{Cosmological perturbations}\label{subsec:pert}

In order to complete the picture of the Early Universe and connect it with current observations, one needs to consider the cosmological perturbations around the given background metric and scalar field. Indeed, cosmological perturbation theory provides predictions for the small inhomogeneities observed in the sky, on the temperature of the Cosmic Microwave Background or the distribution of galaxies in the cosmic web. From expressing the action in an ADM formalism, the symmetries of the background allow to decompose the perturbations into independent scalar, vector and tensor components with respect to how they transform under spatial rotations. One can then find the equations of motion and determine the evolution of the cosmological perturbations. In these equations, the Hubble radius arises playing the role of differentiating the scales in which the modes oscillate (sub-Hubble scales, $k \gg aH$) from the modes which are frozen (super-Hubble scales, $k \ll aH$) and the dynamics is dominated by gravity. The computation of the spectrum of perturbations on super-Hubble scales connects with the high-precision cosmological measures of the Cosmic Microwave Bacground (CMB) and the Large Scale Structures (LSS), which are consistent with such a single-field slow-roll inflationary model. 
Note here that when performing a quantization of the cosmological perturbations, the choice of vacuum state is crucial. When considering the single-field slow-roll inflationary model, the preferred vacuum state is given by the Bunch-Davies state.

The previously introduced single-field slow-roll model of inflation allows one to start from the Einstein-Hilbert action minimally coupled to a scalar field ($\phi$):
\begin{equation}
    S=\frac{M_{\text{Pl}}^2}{2}\int_\M \varepsilon_\M \left[R-(\nabla\phi)^2-2V(\phi)\right],
\end{equation}
where $M_{\text{Pl}}^{-2}=8\pi G$ is the Planck mass, $\varepsilon_{\M}$ is the volume form on $\M$, $R$ the Ricci scalar and $V(\phi)$ the potential. We compute the perturbations around a FLRW metric in the ADM formalism, which slices the spacetime in three-dimensional constant-time hypersurfaces. In this formalism, $\phi$ and the spatial metric $h_{ij}$ are the dynamical fields of the theory, so we choose the following comoving gauge
\begin{equation}
    \delta\phi=0, \quad h_{ij}=a^2 [(1+2\R)\delta_{ij}+\gamma_{ij}], \quad \partial_i \gamma_{ij}=0, \quad \gamma_{ii}=0,
\end{equation}
where $\R$ denotes the metric fluctuation (measuring the spatial curvature of hypersurfaces of constant $\phi$). Expanding the equations in powers of $\R$ to second order one ends up with the following expression for the action~\cite{Baumann:2009ds}
\begin{equation}
\label{adm}
    ^{(2)}S_\R=\frac{M_{\text{Pl}}^2}{2}\int\dd^4x a^3\frac{\dot{\phi}^2}{H^2}[\dot{\R}^2-a^{-2}(\partial \R)^2]=M_{\text{Pl}}^2\int\dd^4x a^3\epsilon_1[\dot{\R}^2-a^{-2}(\partial \R)^2]. 
\end{equation}
Let us perform a change of variables $t\mapsto\eta$ such that $\dd t=a\dd\eta$ and derivatives with respect to the conformal time are denoted with primes. Then 
\begin{equation}
    ^{(2)}S_\R=\frac{M_{\text{Pl}}^2}{2}\int\dd\eta\wedge\dd^3x 2a^2\epsilon_1\left[ (\R')^2 + (\partial\R)^2 \right].
\end{equation}
Let $z^2=2a^2M_{\text{Pl}}^2\epsilon_1$ and let $v=z\R$ be the Mukhanov-Sasaki variable. Then this action is equivalent to
\begin{equation}
\label{actionperturb}
    ^{(2)}S_v=\frac{1}{2}\int\dd\eta\wedge\dd^3x\left[ (v')^2+(\partial v)^2 +\frac{z''}{z}v^2 \right].
\end{equation}
The main interest of the Mukhanov-Sasaki variable relies in its convenient limit in the asymptotic past, where $z \rightarrow 0$. In this limit, the effective mass term vanishes as $z''/z \rightarrow 0$, such that one can quantize the set of canonical variables as a free field in flat spacetime.

\paragraph{Quantization.}
\label{quant}

From the action of Eq. (\ref{actionperturb}), we can perform a quantization of the scalar perturbations.
The momentum density conjugate to $v$ is:
\begin{equation}
    p=\frac{\partial\L_v}{\partial v'}=v',
\end{equation}
and thus the Hamiltonian density:
\begin{equation}
    \H_v=pv'-\L_v=\frac{1}{2}\left[ (v')^2-(\partial v)^2 -\frac{z''}{z}v^2 \right]=\frac{1}{2}\left[ p^2-(\partial v)^2 -\frac{z''}{z}v^2 \right].
\end{equation}
Let us expand the field in the Fourier basis:
\begin{align}
\begin{split}
    &v=\frac{1}{(2\pi)^3}\int\dd^3\boldsymbol{k}v_{\boldsymbol{k}}e^{-i\boldsymbol{k}.\boldsymbol{x}},\\
    &p=\frac{1}{(2\pi)^3}\int\dd^3\boldsymbol{k}p_{\boldsymbol{k}}e^{-i\boldsymbol{k}.\boldsymbol{x}}.
\end{split}
\end{align}
Integrating over a constant time hypersurface yields the Hamiltonian:
\begin{equation}
    H=\frac{1}{2}\int\frac{\dd^3 \boldsymbol{k}}{(2\pi)^3} \left(p_{\boldsymbol{k}}p_{-\boldsymbol{k}}+\omega_{k}^2v_{\boldsymbol{k}}v_{-\boldsymbol{k}}\right),
\end{equation}
where $\omega_{k}^2(\eta)=|\boldsymbol{k}|^2-\frac{z''}{z}=k^2-\frac{(a\sqrt{\epsilon_1})''}{a\sqrt{\epsilon_1}}$. One may notice that the evolution decouples different $k-$sectors so we may consider them separately and denote by $H_{\boldsymbol{k}}$ their respective Hamiltonians.  

Let us consider a case of de Sitter expansion i.e. when $a(\eta)=-1/(H\eta)$ for $\eta \in ] - \infty, 0]$. Then:
\begin{equation}
    \omega^2_{\boldsymbol{k}}(\eta)=k^2-\frac{2}{\eta^2}.
\end{equation}
Defining the creation/annihilation operators for a Bunch-Davies vacuum as
\begin{equation}
    \hat{v}_{\boldsymbol{k}}=\frac{1}{\sqrt{2k}}(\hat{c}_{\boldsymbol{k}}+\hat{c}_{-\boldsymbol{k}}^\dagger)\quad \text{and} \quad \hat{p}_{\boldsymbol{k}}=-i\sqrt{\frac{k}{2}}(\hat{c}_{\boldsymbol{k}}-\hat{c}_{-\boldsymbol{k}}^\dagger),
\end{equation}
gives a quantized Hamiltonian
\begin{equation}
    \hat{H}_{\boldsymbol{k}}=\frac{k}{2}\left[\left(\frac{\omega_{k}^2}{k^2}+1\right)\left(\hat{c}^\dagger_{\boldsymbol{k}}\hat{c}_{\boldsymbol{k}}+\hat{c}_{-\boldsymbol{k}}^\dagger \hat{c}_{-\boldsymbol{k}}+1\right)+\left(\frac{\omega_{k}^2}{k^2}-1\right)\left(\hat{c}_{\boldsymbol{k}}\hat{c}_{-\boldsymbol{k}}+\hat{c}_{-\boldsymbol{k}}^\dagger \hat{c}_{\boldsymbol{k}}^\dagger\right)\right].
\end{equation}
Note that the full Hamiltonian integrates only over the $\mathds{R}^{3+}$ to avoid overcounting. Note also that, under the standard quantization of the ladder operators $c_{\boldsymbol{k}}$ we enacted, the operators $v_{\boldsymbol{k}}$ and $p_{\boldsymbol{k}}$ do not form a canonical pair ($v_{\boldsymbol{k}}$ is actually canonically conjugated to $p_{-\boldsymbol{k}}$). In the following, we will still be allowed to treat our system as if it were a single-mode Gaussian system, because of the symplectic structure clarified in Appendix \ref{2for1}.

\paragraph{Evolution.}

A keen reader should recognize this Hamiltonian as a free oscillator and a squeezer with different $\omega_{k}-$dependent strengths. 
The Heisenberg evolution with this Hamiltonian consists of finding the Boguliubov coefficients $u_{k}, \; w_{k}$:
\begin{equation}
    \begin{pmatrix}
        \hat{c}_{\boldsymbol{k}}(\tau)\\
        \hat{c}_{-\boldsymbol{k}}^\dagger(\tau)
    \end{pmatrix}=S_{c_v}\begin{pmatrix}
        \hat{c}_{\boldsymbol{k}}(\tau_\text{in})\\
        \hat{c}_{-\boldsymbol{k}}^\dagger(\tau_\text{in})
    \end{pmatrix}, \qquad \text{with} \qquad S_{c_v} \equiv \begin{pmatrix}
    u_{k}(\tau) &  w_{k}(\tau)\\
    w^*_{k}(\tau) &  u^*_{k}(\tau)
\end{pmatrix}.
\end{equation}
Note that the non-zero commutators are
\begin{align}
    &[\hat{c}_{\boldsymbol{k}}^\dagger \hat{c}_{\boldsymbol{k}},\hat{c}_{\boldsymbol{q}}]=-\hat{c}_k(2\pi)^3\delta^3(\boldsymbol{k}-\boldsymbol{q}),\qquad [\hat{c}_{\boldsymbol{k}}^\dagger \hat{c}_{\boldsymbol{k}},\hat{c}_{\boldsymbol{q}}^\dagger]=\hat{c}_k^\dagger(2\pi)^3\delta^3(\boldsymbol{k}-\boldsymbol{q}),
\end{align}
together with 
\begin{align}
    &[\hat{c}_{-{\boldsymbol{k}}}^\dagger \hat{c}_{\boldsymbol{k}}^\dagger,\hat{c}_{\boldsymbol{q}}]=-\hat{c}_{-{\boldsymbol{k}}}^\dagger(2\pi)^3\delta^3(\boldsymbol{k}-\boldsymbol{q})-\hat{c}_{\boldsymbol{k}}^\dagger(2\pi)^3\delta^3(-\boldsymbol{k}-\boldsymbol{q}),\\
    &[\hat{c}_{{\boldsymbol{k}}} \hat{c}_{-{\boldsymbol{k}}},\hat{c}_{\boldsymbol{q}}^\dagger]=\hat{c}_{-{\boldsymbol{k}}}(2\pi)^3\delta^3(\boldsymbol{k}-\boldsymbol{q})+\hat{c}_k(2\pi)^3\delta^3(-\boldsymbol{k}-\boldsymbol{q}).
\end{align}
Thus, the Heisenberg equations of motion are:
\begin{equation}
   \frac{\dd}{\dd\eta} \begin{pmatrix}
        \hat{c}_{\boldsymbol{k}}\\
        \hat{c}_{\boldsymbol{k}}^\dagger\\
        \hat{c}_{-\boldsymbol{k}}\\
        \hat{c}_{-\boldsymbol{k}}^\dagger
    \end{pmatrix}=i\frac{k}{2}\begin{pmatrix}
        -\left(\frac{\omega_{k}^2}{k^2}+1\right) & 0 & 0 & -\left(\frac{\omega_{k}^2}{k^2}-1\right)\\
        0 & \left(\frac{\omega_{k}^2}{k^2}+1\right) & \left(\frac{\omega_{k}^2}{k^2}-1\right) & 0\\
        0 & -\left(\frac{\omega_{k}^2}{k^2}-1\right) & -\left(\frac{\omega_{k}^2}{k^2}+1\right) & 0\\
        \left(\frac{\omega_{k}^2}{k^2}-1\right) & 0 & 0 & \left(\frac{\omega_{k}^2}{k^2}+1\right)
    \end{pmatrix}\begin{pmatrix}
        \hat{c}_{\boldsymbol{k}}\\
        \hat{c}_{\boldsymbol{k}}^\dagger\\
        \hat{c}_{-\boldsymbol{k}}\\
        \hat{c}_{-\boldsymbol{k}}^\dagger
    \end{pmatrix}. 
\end{equation}
This is written somewhat redundantly and can be compactly recast as:
\begin{equation}
\label{eq:eom}
   \frac{\dd}{\dd \eta} \begin{pmatrix}
        \hat{c}_{\boldsymbol{k}}\\
        \hat{c}_{-\boldsymbol{k}}^\dagger
    \end{pmatrix}=i\frac{k}{4}\begin{pmatrix}
        -\left(\frac{\omega_{k}^2}{k^2}+1\right) & -\left(\frac{\omega_{k}^2}{k^2}-1\right)\\
        \left(\frac{\omega_{k}^2}{k^2}-1\right) & \left(\frac{\omega_{k}^2}{k^2}+1\right)
    \end{pmatrix}\begin{pmatrix}
        \hat{c}_{\boldsymbol{k}}\\
        \hat{c}_{-\boldsymbol{k}}^\dagger
    \end{pmatrix}.
\end{equation}
The Boguliubov coefficients should satisfy the same differential equation. Let us define a new function $\phi_k=u_k+w_k^*$. Notice that $\phi_k'=u_k'+{w_k^*}'= -ik(u_k-w_k^*)$, such that
\begin{align}
   \phi_k''=-ik(u_k'-{w_k^*}')= -\omega_{k}^2(u_k+w_k^*)=-\omega_{k}^2\phi_k.
\end{align}
For the $\omega_{k}$ assumed earlier this yields a differential equation
\begin{equation}
    \phi''_k+\left(k^2-\frac{2}{\eta^2}\right)\phi_k=0.
\end{equation}
The normal mode is then
\begin{equation}
    \phi_k=\left(1-\frac{i}{k\eta}\right)e^{-ik\eta}.
\end{equation}
Note that the differential equation admits also an incoming wave solution. However, we assume boundary conditions with outgoing waves only (a.k.a. the Bunch-Davies vacuum). Using the form of the derivative we can find the Boguliubov coefficients which are given by
\begin{equation}
    u_{k}=\left(1-\frac{i}{k\eta}-\frac{1}{2k^2\eta^2}\right)e^{-ik\eta}, \quad w_{k}^*=\frac{1}{2k^2\eta^2}e^{-ik\eta}.
\end{equation}
It is easy to verify that they satisfy the Wronskian condition $|u_k|^2-|w_k|^2=1$, thus the evolution is symplectic.
The evolution is thus given by a symplectic matrix of the form:
\begin{equation}\label{symp}
    S_{c_v}=\begin{pmatrix}
        \left(1-\frac{i}{k\eta}-\frac{1}{2k^2\eta^2}\right)e^{-ik\eta} & \frac{1}{2k^2\eta^2}e^{ik\eta}\\
        \frac{1}{2k^2\eta^2}e^{-ik\eta} & \left(1+\frac{i}{k\eta}-\frac{1}{2k^2\eta^2}\right)e^{ik\eta}
    \end{pmatrix}.
\end{equation}

\paragraph{Covariance} The covariance matrix is then given by
\begin{equation}
    \sigma_{c_v}\equiv S_{c_v}S^\dagger_{c_v}=\begin{pmatrix}
        1+\frac{1}{2 k^4\eta^4} & \frac{-1+2k\eta(k\eta-i)}{2 k^4 \eta^4}\\
        \frac{-1+2k\eta(k\eta+i)}{2 k^4 \eta^4} &  1+\frac{1}{2 k^4\eta^4} 
    \end{pmatrix}.
\end{equation}
It is often more convenient to work in the quadrature basis. Recall:
\begin{equation}
    \begin{pmatrix}
        \hat{v}_{\boldsymbol{k}}\\
        \hat{p}_{\boldsymbol{k}}
    \end{pmatrix}=\begin{pmatrix}
        \frac{1}{\sqrt{2k}} & \frac{1}{\sqrt{2k}}\\
        -i\sqrt{\frac{k}{2}} & i\sqrt{\frac{k}{2}}
    \end{pmatrix}\begin{pmatrix}
        \hat{c}_{\boldsymbol{k}}\\
        \hat{c}^\dagger_{-\boldsymbol{k}}
    \end{pmatrix}\coloneqq U \begin{pmatrix}
        \hat{c}_{\boldsymbol{k}}\\
        \hat{c}^\dagger_{-\boldsymbol{k}}
    \end{pmatrix}.
\end{equation}
Changing basis yields:
\begin{equation}
    \sigma_v=U\sigma_{c_v}U^\dagger=\begin{pmatrix}
        \frac{1}{k}\left( 1+ \frac{1}{k^2\eta^2} \right) & -\frac{1}{k^3\eta^3}\\
        -\frac{1}{k^3\eta^3} & k\left( 1 -\frac{1}{k^2 \eta^2} + \frac{1}{k^4\eta^4}\right)
    \end{pmatrix}.
\end{equation}

\begin{figure}[tbp]
    \centering
    \includegraphics[width=0.7\textwidth]{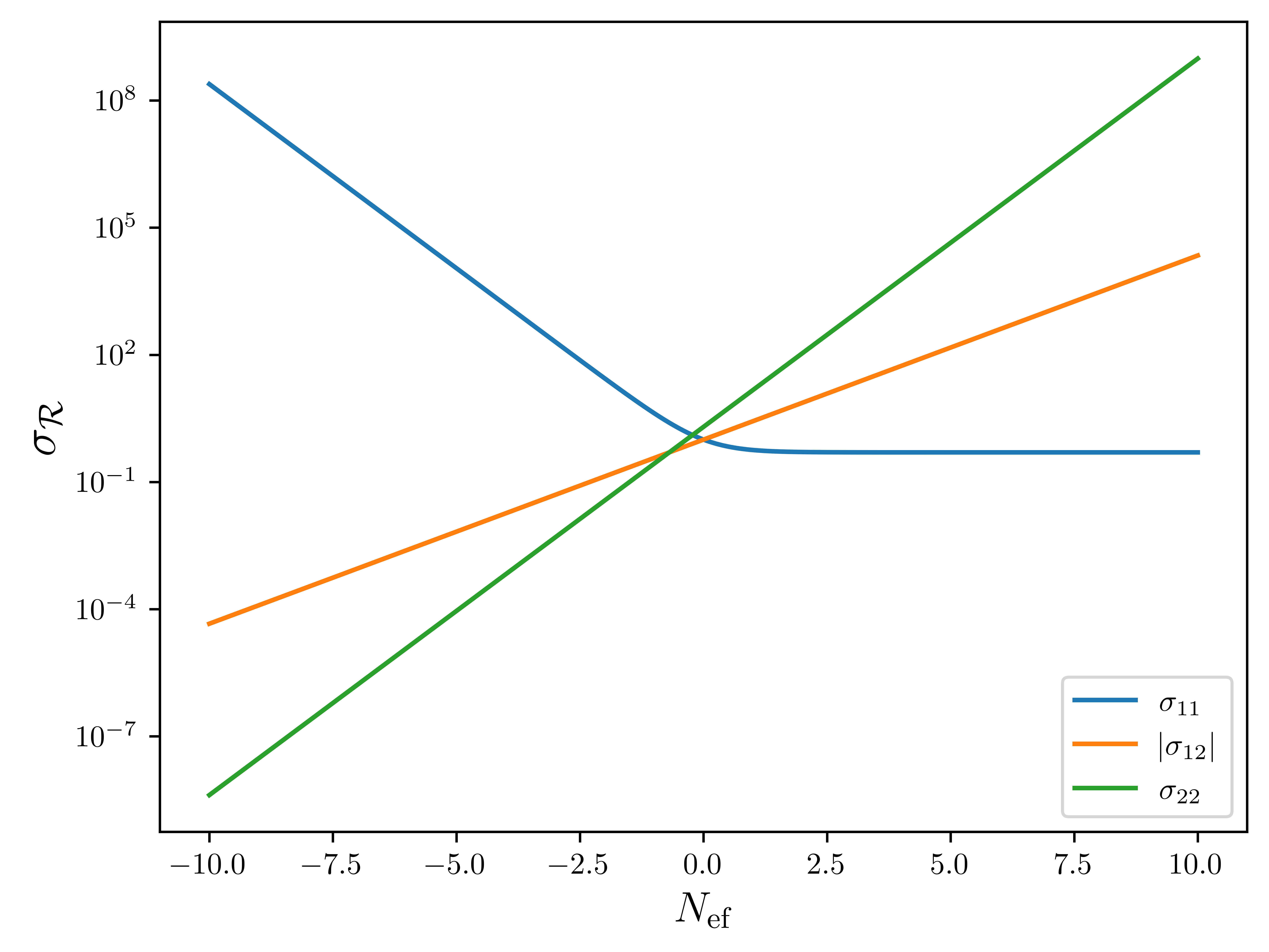}
\caption{Entries of the covariance matrix $\sigma_\R$ as a function of the number of \textit{e}-folds $N_{\mathrm{ef}} \equiv - \log (- k \eta)$}. The first entry $\sigma_{11}$ (\textit{Blue} curve) freezes at late time, on super-Hubble scales.
    \label{fig:cov}
\end{figure}

To get back to curvature perturbations, we use the classical canonical transformation:
\begin{equation}\label{class}
    \begin{pmatrix}
        \hat{v}_{\boldsymbol{k}}\\
        \hat{p}_{\boldsymbol{k}}
    \end{pmatrix}=\begin{pmatrix}
        z & 0\\
        z' & 1/z
    \end{pmatrix}\begin{pmatrix}
        \hat{\R}_{\boldsymbol{k}}\\
        \hat{\P}_{\boldsymbol{k}}
    \end{pmatrix}\coloneqq M \begin{pmatrix}
        \hat{\R}_{\boldsymbol{k}}\\
        \hat{\P}_{\boldsymbol{k}}
    \end{pmatrix},
\end{equation}
where we remind that $z = aM_{\text{Pl}} \sqrt{2\epsilon_1}$ and $a=-1/(H\eta)$. The resulting covariance matrix is
\begin{equation}\label{sigmar}
    \sigma_\R = M^{-1}\sigma_v (M^{-1})^T=\begin{pmatrix}
        \frac{H^2}{2 k^3M_{\text{Pl}}^2 \epsilon_1}(1+k^2\eta^2) & \frac{1}{k \eta}\\
        \frac{1}{k \eta} & \frac{2 kM_{\text{Pl}}^2\epsilon_1}{H^2 \eta^2}
    \end{pmatrix}.
\end{equation}
The entries of the covariance matrix $\sigma_\R$ are given in Fig. \ref{fig:cov}. This set of variables is particularly appreciated by cosmologists due to the fact the first entry of the covariance matrix (\textit{Blue} curve) freezes at late time, on super-Hubble scale where $- k \eta \ll 1$.

\begin{figure}[htbp]
    \centering
    \begin{minipage}{0.48\textwidth}
        \centering
        \includegraphics[width=\textwidth]{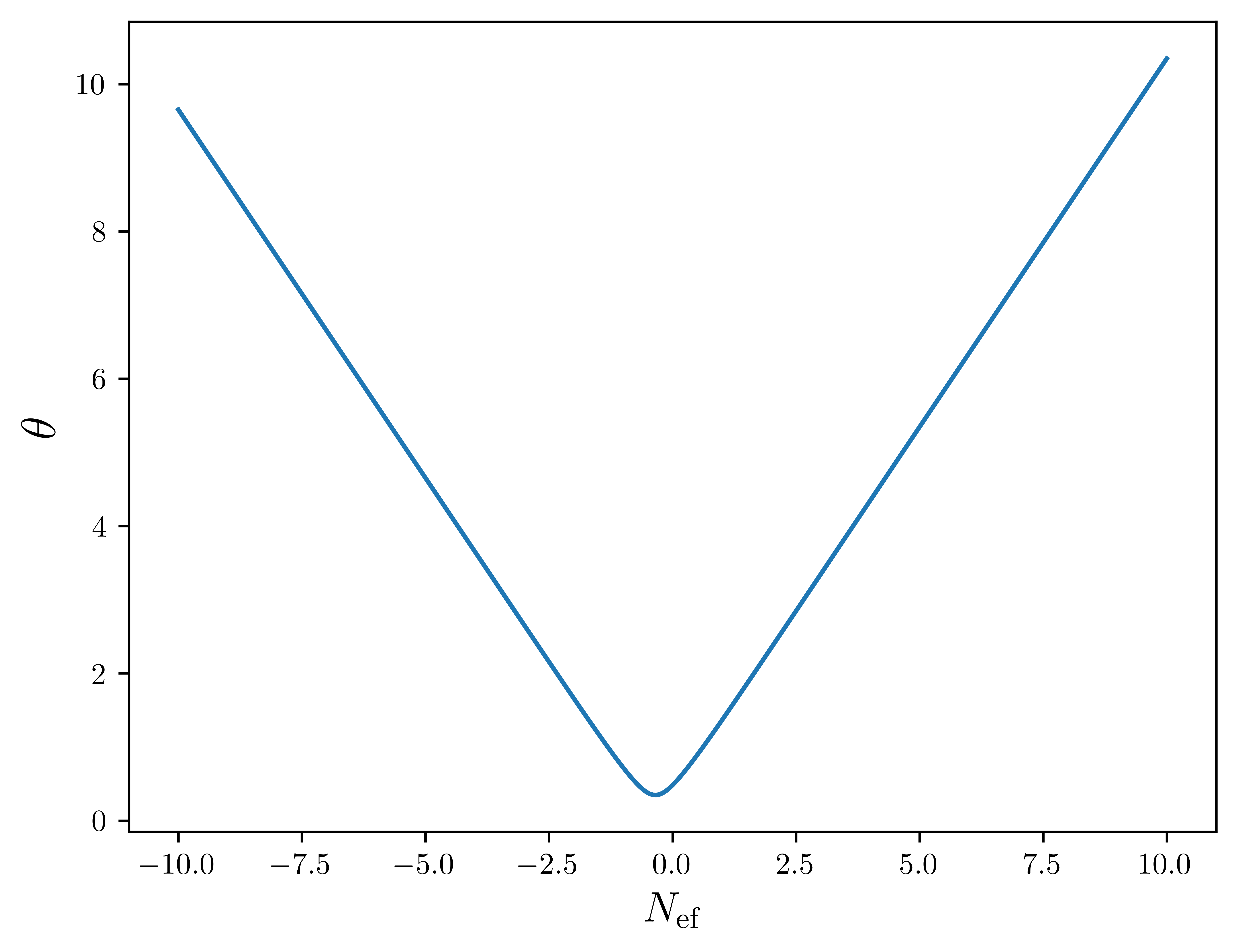}
    \end{minipage}
    \hfill
    \begin{minipage}{0.48\textwidth}
        \centering
        \includegraphics[width=\textwidth]{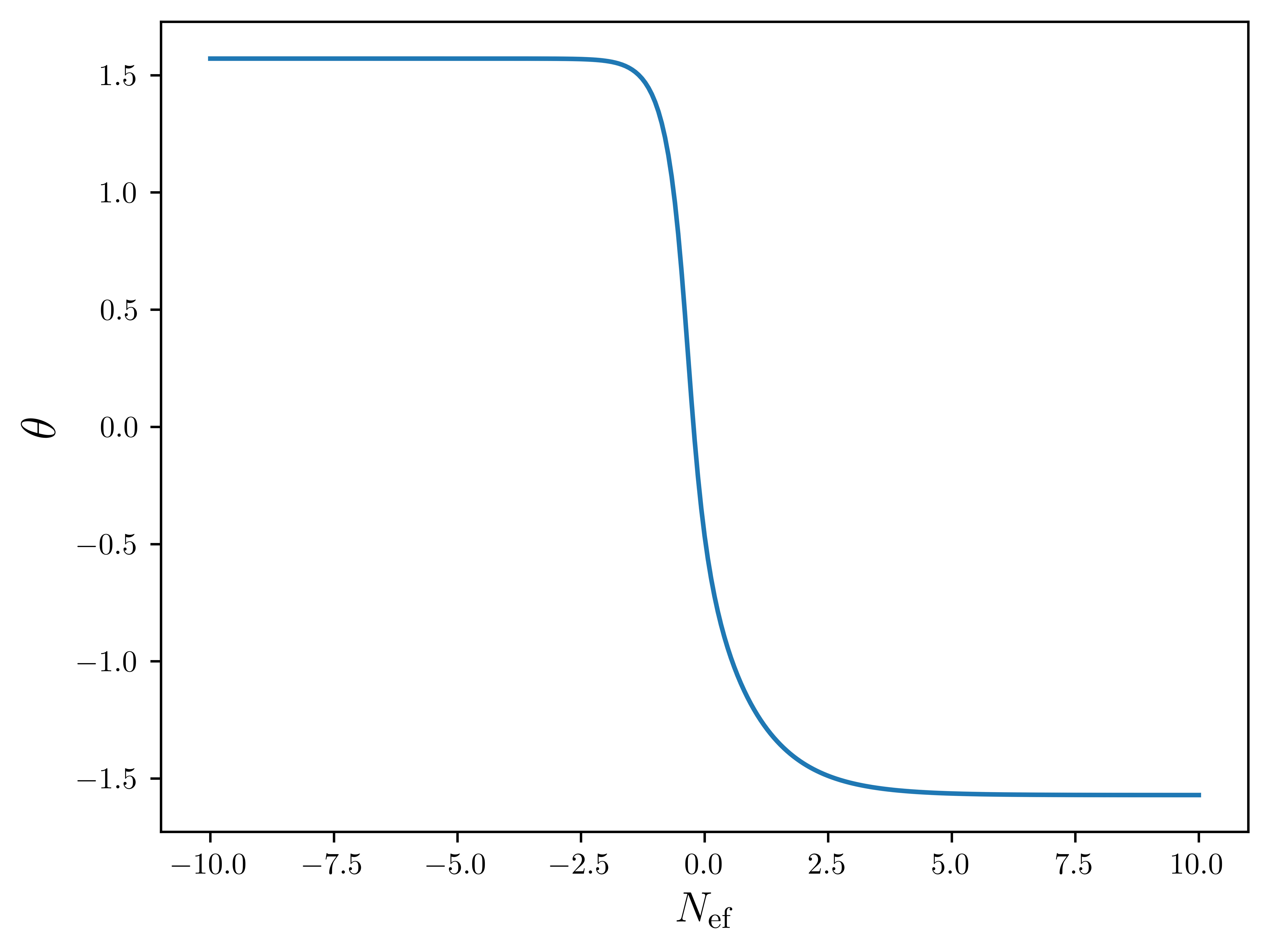}
    \end{minipage}
    \caption{Squeezing parameters as a function of the number of \textit{e}-folds $N_{\mathrm{ef}} \equiv - \log (- k \eta)$. \textit{Left:} Squeezing parameter $r$. \textit{Right:} Squeezing angle $\theta$.}
    \label{fig:squeezing}
\end{figure}

\paragraph{Squeezing.} Let us consider a parametrization of this Gaussian state by angle $\theta$ and squeezing parameter $r$ such that
\begin{equation}
    \sigma_\R=S(\theta,r)S^T(\theta,r),
\end{equation}
where
\begin{equation}
    S(\theta,r)=R(\theta/2)\begin{pmatrix}
        e^{r} & 0\\
        0 & e^{-r}
    \end{pmatrix}R(-\theta/2).
\end{equation}
Then
\begin{equation}\label{eq:covmat}
    \sigma_\R=\begin{pmatrix}
        \cosh(2r)+\cos(\theta)\sinh(2r) & -\sin(\theta)\sinh(2r)\\
        -\sin(\theta)\sinh(2r) &  \cosh(2r)-\cos(\theta)\sinh(2r)
    \end{pmatrix}.
\end{equation}
Note that $\Tr(\sigma_\R)=2\cosh(2r)$ and thus
\begin{equation}\label{eq:sqref1}
    r=\frac{1}{2}\cosh^{-1}\left[ \frac{kM_{\text{Pl}}^2\epsilon_1}{H^2 \eta^2} + \frac{H^2}{4 k^3 M_{\text{Pl}}^2\epsilon_1}(1+k^2\eta^2) \right] = \begin{dcases}
 \frac{1}{2}\log \frac{H^2 \eta^2}{2 k \Mpl^2 \epsilon_1}, & -k \eta \gg 1 ,\\
  \frac{1}{2}\log \frac{2 k \Mpl^2 \epsilon_1}{H^2 \eta^2}, & -k \eta \ll 1 .
\end{dcases}.
\end{equation}
We can also find the angle as: 
\begin{equation}\label{eq:sqref2}
    \theta = \tan^{-1}\left[ \frac{4 H^2 k^4 \eta M_{\text{Pl}}^2 \epsilon_1}{4 k^6M_{\text{Pl}}^4\epsilon_1^2-H^4 k^2 \eta^2 (1+k^2\eta^2)} \right]=  \begin{dcases}
 \pi + \frac{4 \Mpl^2 \epsilon_1}{H^2 \eta^3} , & -k \eta \gg 1 ,\\
  - \frac{H^2 \eta}{k^2 \Mpl^2 \epsilon_1}, & -k \eta \ll 1 .
\end{dcases}.
\end{equation}
The time dependence of these parameters is given in Fig.~\ref{fig:squeezing} for a given mode $k$. The state is highly squeezed in the asymptotic past along one direction in the phase space, then highly squeezed in the asymptotic future in the orthogonal direction.

\section{Quantum Fisher information}\label{sec:QFI}

Now that we have formulated the state of interest in terms of squeezing parameters, we turn our attention to the parameter estimation one can expect with this state.

\subsection{Gaussian parameter estimation}

The {\em classical} Fisher information $I_{\mathrm{cl}}$ yields, through the Cram\'er-Rao bound, an absolute lower bound to the variance of a parameter $\vartheta$ estimated upon sampling a probability distribution: $(\Delta \vartheta)^2 \ge\frac{1}{N I_{\mathrm{cl}}}$, where $N$ is the number of independent measurements taken. Once a POVM is set, measuring a quantum system boils down to sampling a classical distribution, to which the Cram\'er-Rao bound applies; the {\em quantum} Fisher information $I$ is then a further optimisation over all possible POVMs, and yields the quantum Cram\'er-Rao bound, yielding the ultimate lower bound to the variance of an estimated parameter encoded in a quantum state, as per \cite{holevo73}
\begin{equation}\label{eq:estim}
   (\Delta \vartheta)^2 \ge \frac{1}{N I} ,
\end{equation}
with
\begin{equation}
   I = {\rm Tr}[\rho L^2_{\vartheta}] ,
\end{equation}
where the symmetric logarithmic derivative (SLD) $L_{\vartheta}$ is a self-adjoint operator such that 
\begin{equation}
2\frac{{\rm d}\rho}{{\rm d}\vartheta} = 2\rho_{,\vartheta} = L_{\vartheta}\rho+\rho L_{\vartheta}. 
\end{equation}
Henceforth a prime $_{,\vartheta}$ will denote derivatives with respect to the parameter in question $\vartheta$.

For a pure state $\rho=\ketbra{\psi}$, which is our case, $\rho=\rho^2$ and therefore $L_\vartheta=2\rho_{,\vartheta}$, as
\begin{equation}
    \rho_{,\vartheta}=(\rho^2)_{,\vartheta}=\rho_{,\vartheta}\rho+\rho\rho_{,\vartheta} .
\end{equation}
The associated quantum Fisher information is thus
\begin{align}
\begin{split}
    I = \Tr[\rho L^2_\vartheta]&=4\Tr\left[\ketbra{\psi}\left(\ket{\psi_{,\vartheta}}\bra{\psi}+\ket{\psi}\bra{\psi_{,\vartheta}}\right)^2\right]\\
    &=4\left(\bra{\psi}{\psi_{,\vartheta}}\rangle^2+|\bra{\psi}{\psi_{,\vartheta}}\rangle|^2+||\psi_{,\vartheta}||^2+\bra{\psi_{,\vartheta}}{\psi}\rangle^2\right).
\end{split}
\end{align}
Notice now that the pure state at hand can always be written as the Hamiltonian evolution (which we expressed in symplectic form) of an initial state $\ket{0}$: $\ket{\psi}={\rm e}^{-i H}\ket{0}$. Hence, 
\begin{align}\label{eq:proof1}
    \bra{\psi_{,\vartheta}}{\psi}\rangle= i\bra{\psi}\partial_\vartheta H\ket{\psi}
\end{align}
must be purely imaginary. Then, letting $\bra{\psi_{,\vartheta}}{\psi}\rangle=Ze^{i\phi}$ for $Z>0$ and $\cos(\phi)=0$ yields
\begin{align}
    |\bra{\psi}{\psi_{,\vartheta}}\rangle|^2+\bra{\psi}{\psi_{,\vartheta}}\rangle^2+\bra{\psi_{,\vartheta}}{\psi}\rangle^2=&Z^2[1+2\cos(2\phi)]  \nonumber \\
    =&Z^2[4\cos^2(\phi)-1]=-Z^2=-|\bra{\psi}{\psi_{,\vartheta}}\rangle|^2.
\end{align}
Thus
\begin{equation}\label{eq:proof2}
    I=4\left(||\psi_{,\vartheta}||^2 -|\bra{\psi}{\psi_{,\vartheta}}\rangle|^2 \right).
\end{equation}
Notice that this is the squared norm of the vector $\ket{\psi_{,\vartheta}}-\bra{\psi}{\psi_{,\vartheta}}\rangle\ket{\psi}$, orthogonal to $\ket{\psi}$ in the plane spanned by $\ket{\psi}$ and $\ket{\psi_{,\vartheta}}$.

Another expedient expression for $I$ may be found by noticing
\begin{align}
    &\Tr[\rho\rho_{,\vartheta}]=2\Re\bra{\psi_{,\vartheta}}{\psi}\rangle ,\\
    &\Tr[\rho_{,\vartheta}\rho_{,\vartheta}]=2||\psi_{,\vartheta}||^2+\bra{\psi}{\psi_{,\vartheta}}\rangle^2+\bra{\psi_{,\vartheta}}{\psi}\rangle^2,
\end{align}
and thus
$I=2(\Tr[\rho_{,\vartheta}\rho_{,\vartheta}]-\Tr[\rho\rho_{,\vartheta}]^2)$.
But since $\Re\braket{\psi_{,\vartheta}|\psi}=0$, one has simply
\begin{equation}\label{pureqfi}
    I=2\Tr[\rho_{,\vartheta}\rho_{,\vartheta}].
\end{equation}

\subsection{Fisher information from the characteristic function}

We can now use Eq.~(\ref{pureqfi}) to readily obtain an expression for the QFI of the pure Gaussian states we are dealing with. Since displacements form a continuous basis in the space of operators, the state of a Gaussian state with a covariance matrix $\sigma$ and null first moments may be written as \cite{serafini2024}
\begin{equation}\label{eq:dispop}
    \rho=\frac{1}{2\pi}\int\dd^2 \Tilde{r}e^{-\frac{1}{4}\sigma_{ab}\Tilde{r}^a\Tilde{r}^b}\hat{D}_{\Omega^T\Tilde{\boldsymbol{r}}},
\end{equation}
where $\hat{D}_{\boldsymbol{r}} = {\rm e}^{i \Omega_{jk}r_j\hat{r}_k}$ is the displacement operator. In general, this is the Fourier-Weyl relation linking the characteristic function to its quantum state.

The derivative of the density matrix $\rho$ can then be evaluated directly at the level of the characteristic function
\begin{equation}
    \rho_{,\vartheta} = \frac{1}{2\pi}\int\dd^2 \Tilde{r}\left(-\frac{1}{4} \sigma_{jk,\vartheta}\Tilde{r}^j\Tilde{r}^k\right)e^{-\frac{1}{4}\sigma_{ab}\Tilde{r}^a\Tilde{r}^b}\hat{D}_{\Omega^T\Tilde{\boldsymbol{r}}}. 
\end{equation}
Hence, using the orthogonality of displacement operators under the Hilbert-Schmidt inner product, one gets
\begin{equation} \label{chara}
    \Tr[\rho_{,\vartheta}\rho_{,\vartheta}]=\frac{1}{2\pi}\frac{1}{16}\int\dd^2 \Tilde{r}\left(\sigma_{jk,\vartheta}\Tilde{r}^j\Tilde{r}^k\right)^2e^{-\frac{1}{2}\sigma_{ab}\Tilde{r}_1^a\Tilde{r}_1^b}.
\end{equation}
This Gaussian integral is easily evaluated as a simple instance of Isserlis' theorem (the classical counterpart to Wick's, whereby higher order moments of Gaussian distributions are the sum of products of the second-order moments of all possibly partitions), yielding the (already known \cite{Pinel12,Jiang14}) formula for the QFI of pure Gaussian states with null first moments:
\begin{equation}\label{isserlis}
I = 2{\rm Tr}[\rho_{,\vartheta}^2] = \frac14 {\rm Tr}[\sigma^{-1}\sigma_{,\vartheta}\sigma^{-1}\sigma_{,\vartheta}] \; .
\end{equation}

A specific case of states we are interested in have the dependence on the parameter only on the main diagonal. Without loss of generality, a pure Gaussian state's covariance matrix may be written as $\sigma=S S^T$ for 
symplectic $S$. 
In the one-mode case, the most general symplectic can be Euler-decomposed to obtain
\begin{align}
    \sigma=R(\theta/2)\begin{pmatrix}
        e^{2r} & 0\\
        0 & e^{-2r}
    \end{pmatrix}R(-\theta/2),
\end{align}
where $R(\theta/2)$ is a two-dimensional rotation by the angle $\theta/2$.
Thus, 
\begin{equation}
    \sigma_{,\vartheta} = \begin{pmatrix}
        \sigma_{11,\vartheta} 
& \sigma_{12,\vartheta} 
\\
\sigma_{12,\vartheta} 
& \sigma_{22,\vartheta} 
    \end{pmatrix},
\end{equation}
with 
\begin{align}
    \sigma_{11,\vartheta} &= 2 \left[\cos(\theta) \cosh(2r) + \sinh(2r)\right] r_{,\vartheta} - \sin(\theta) \sinh(2r) \theta_{,\vartheta} ,\\
    \sigma_{12,\vartheta}  &= 2 \left[\cos(\theta) \cosh(2r) + \sinh(2r)\right] r_{,\vartheta} - \sin(\theta) \sinh(2r) \theta_{,\vartheta} ,\\
    \sigma_{22,\vartheta} &= 2 \left[-\cos(\theta) \cosh(2r) + \sinh(2r)\right] r_{,\vartheta} + \sin(\theta) \sinh(2r) \theta_{,\vartheta}.
\end{align}

By noticing that $R_{,\vartheta}= \Omega R \theta_{,\vartheta}/2$, it is easy to see that the previous formula is equivalent to
\begin{equation}
    \sigma_{,\vartheta} = R(\theta/2) \left[2r_{,\vartheta}Z\sigma_z-\sinh(2r)\sigma_x\theta_{,\vartheta}\right] R(-\theta/2) ,
\end{equation}
where $\sigma_z$ and $\sigma_x$ are the customary Pauli matrices and $Z={\rm diag}({\rm e}^{2r},{\rm e}^{-2r})$.
Therefore, 
\begin{equation}
\sigma^{-1}\sigma_{,\vartheta}=R(\theta/2) \left[2r_{,\vartheta}\sigma_z-\sinh(2r)Z^{-1}\sigma_x\theta_{,\vartheta}\right] R(-\theta/2).
\end{equation}
Then, under the condition of parameter-independence of the off-diagonal elements
\begin{equation}
    2 \cosh(2r) \sin(\theta) r_{,\vartheta} 
+ \cos(\theta) \sinh(2r) \theta_{,\vartheta} =0 .
\end{equation}
Eq.~(\ref{isserlis}) yields
\begin{equation}\label{eq:QFIfin}
    I=\left[ (1 + \cosh(4r)) \sec^2(\theta) - 2 \sinh^2(2r) \right] r_{,\vartheta}^2=2\left[\cosh^2(2r)\tan^2(\theta)+1\right]r_{,\vartheta}^2 .
\end{equation}
This results provide a simple relation between the squeezing parameters and the associated QFI. An alternative derivation of this formula can be found in Appendix \ref{app:app2} using the Fock's space representation of the state. 
Notice that, in the case at hand, we should apply this result to the covariance matrix $\sigma_{\R}$ of Eq.~(\ref{sigmar}), which refers to the quadratures 
\begin{align}\label{qfigen}
	\hat{\R}_{\bf k}&=\frac{\hat{c}_{\bf k}}{z\sqrt{2k}}+\frac{\hat{c}^{\dag}_{-\bf k}}{z\sqrt{2k}}, \qquad \hat{\P}_{\bf k}=\left(-\frac{z'}{\sqrt{2k}}-iz\sqrt{\frac{k}{2}}\right)\hat{c}_{\bf k}+\left(-\frac{z'}{\sqrt{2k}}+iz\sqrt{\frac{k}{2}}\right)\hat{c}^{\dag}_{-\bf k}.
\end{align} 

Here, one should bear in mind that access to the pair $\hat{\R}_{-\bf k}$ and $\hat{\P}_{-\bf k}$, which complete the two-mode canonical basis (as per Appendix \ref{2for1}) would allow one to double the QFI, as apparent from the integral formula (\ref{chara}), where the integral occur independently on each pair of coupled phase-space variables.

Eq.~(\ref{eq:QFIfin}), being general for our class of states, shows that the QFI has an exponential dependence on the squeezing parameter, except for cases where the derivative $r_{,\vartheta}$ goes to zero, where cancellations may occur. In our specific case, as we are about to see, the QFI diverges asymptotically at the end of inflation, but not at $\eta\rightarrow -\infty$: in fact, although $r$ diverges linearly with $\log(-k\eta)$ and 
$\theta\rightarrow \mp\pi/2$, so that $\tan^2\theta$ also diverges, as apparent from Eqs.~(\ref{eq:sqref1}) and (\ref{eq:sqref2}) and shown in Fig.~\ref{fig:squeezing}, the derivative $r_{,\vartheta}^2$ cancels the divergence at $\eta\rightarrow-\infty$. Heuristically, such asymtpotic divergences of the QFI are related to the increased, phase-dependent sensitivity of our states at large squeezing parameters, for which the vanishing noise on the squeezed quadrature can be transferred to the parameter of interest through the optimal POVM. Neighbouring states (in terms of the parameter $\vartheta$) may, however, no longer be distinguished if the derivative of the squeezing parameter goes to zero as well.

Albeit our expression for the quantum Fisher information was re-derived from scratch, under the specific condition of independence of the off-diagonal CM elements, it is worthwhile noting here that, besides the general approaches that may be found in \cite{Pinel12,monras13,Jiang14,serafini2024}, the estimation of Gaussian parameters has a longstanding tradition in quantum optics, featuring systematic studies on the estimation of squeezing parameters \cite{milburn94,chiribella06,rigovacca17}, on the estimation enhancements allowed by squeezing \cite{gaiba09,genoni09}, on multimode scenarios \cite{safranek15}, on the estimation of general Gaussian unitaries \cite{safranek16}, as well as on multiparameter estimation \cite{bressanini24,chang25}. 

\section{Application to inflationary squeezed states}\label{sec:appinfl}

It is now time to combine the dynamics of scalar perturbations in the early universe (Sec.~\ref{chap:scalar}) with the quantum Fisher information framework used in metrology (Sec.~\ref{sec:QFI}). A key characteristic of inflationary dynamics is that it places cosmological inhomogeneities in a highly squeezed state on scales that are observable today~\cite{Grishchuk:1990bj, Albrecht:1992kf}. This feature is dual in nature: it exhibits strongly quantum aspects — such as the vacuum pair creation process responsible for generating inhomogeneities — while also appearing highly classical in other respects. For instance, the practical impossibility of measuring correlators of the conjugate momenta prevents the direct observation of genuine quantum correlations in the sky~\cite{Grain:2019vnq, Colas:2021llj}.

As we will see below, the comparison between the optimal inference captured by the quantum Fisher information and its classical counterpart — once a POVM is specified — reflects this ambiguity. The key distinction between quantum and classical Fisher information lies in the minimization over POVMs. If we could physically perform such a minimization, which requires access to the expectation values of any quantum operator at the end of inflation, we would be able to infer a physical parameter with maximum precision.
However, the highly efficient squeezing mechanism during inflation prevents this. In practice, we are constrained to measuring only the correlators of the position operator, $\langle \hat{\mathcal{R}}^n\rangle$. In particular, the moments of the conjugate momentum, $\langle \hat{\mathcal{P}}^n\rangle$, are observationally inaccessible, as they require access to the so-called decaying mode — associated with the velocity field $\dot{\mathcal{R}}$ — which becomes exponentially suppressed with the number of \textit{e}-folds, $N_{\mathrm{ef}}$, that a given mode spends outside the horizon.\footnote{\label{foot:nef}This statement might appear counterintuitive in light of the late-time growth of the covariance entries $\sigma_{12}$ and $\sigma_{22}$ shown in Fig.~\ref{fig:cov}. However, this growth reflects the unobservability of the conjugate momentum. Indeed, the conversion factor $\dot{\mathcal{R}} \propto a^{-3} \,\mathcal{P}$ implies that measuring $\sigma_{12} \propto a$ would require access to an observable scaling as $a^{-2} \sim \exp(-2N_{\mathrm{ef}})$, and similarly, $\sigma_{22} \propto a^2$ would require access to a quantity scaling as $a^{-4} \sim \exp(-4N_{\mathrm{ef}})$. These exponential suppressions with the number of \textit{\textit{e}-folds} spent outside the horizon — typically $N_{\mathrm{ef}} \simeq 50$ for CMB modes — render such measurements practically impossible.} For CMB-observable modes, $N_{\mathrm{ef}} \simeq 50$ under a standard reheating scenario, which effectively precludes any experimental access to these quantities. By comparing the quantum Fisher information with the classical Fisher information obtained by specifying a POVM along the position direction, we can assess whether the measurement imposed by nature is nearly optimal or highly suboptimal.

In practice, these methods require to pick up a parameter one wants to infer. For the purpose of this work, we focussed on the \textit{tensor-to-scalar ratio r}, computed at leading order in the slow-roll expansion
 \begin{align}
	r \equiv \frac{2 \mathrm{\sigma}_{\gamma,11}}{\mathrm{\sigma}_{\mathcal{R},11}} = 16 \epsilon_1.
\end{align}
Equivalently, this corresponds to estimating the precision with which a measurement can infer the first slow-roll parameter, $\epsilon_1$. At leading order in slow-roll, $\epsilon_1$ appears only in the covariance matrix of the scalar sector, which simplifies the derivation of the quantum Fisher information. Ultimately, this computation informs us whether we are fortunate enough that our measurements provide meaningful constraints on the tensor-to-scalar ratio, or whether we face the unfortunate scenario in which the only accessible observables have poor sensitivity to the parameter we seek to infer.

In Sec.~\ref{subsec:QFI}, we derive the quantum Fisher information associated to the inference of the first slow-roll parameter using the two-mode squeezed state in which are placed the cosmological inhomogeneities. Then, in Sec.~\ref{subsec:optivshomo}, we compare this quantity to the classical Fisher information obtained once imposing a homodyne measurement along the position axis of the quadratures. 

\subsection{Quantum Fisher information}\label{subsec:QFI}

\begin{figure}[htbp]
    \centering
    \begin{minipage}{0.48\textwidth}
        \centering
        \includegraphics[width=\textwidth]{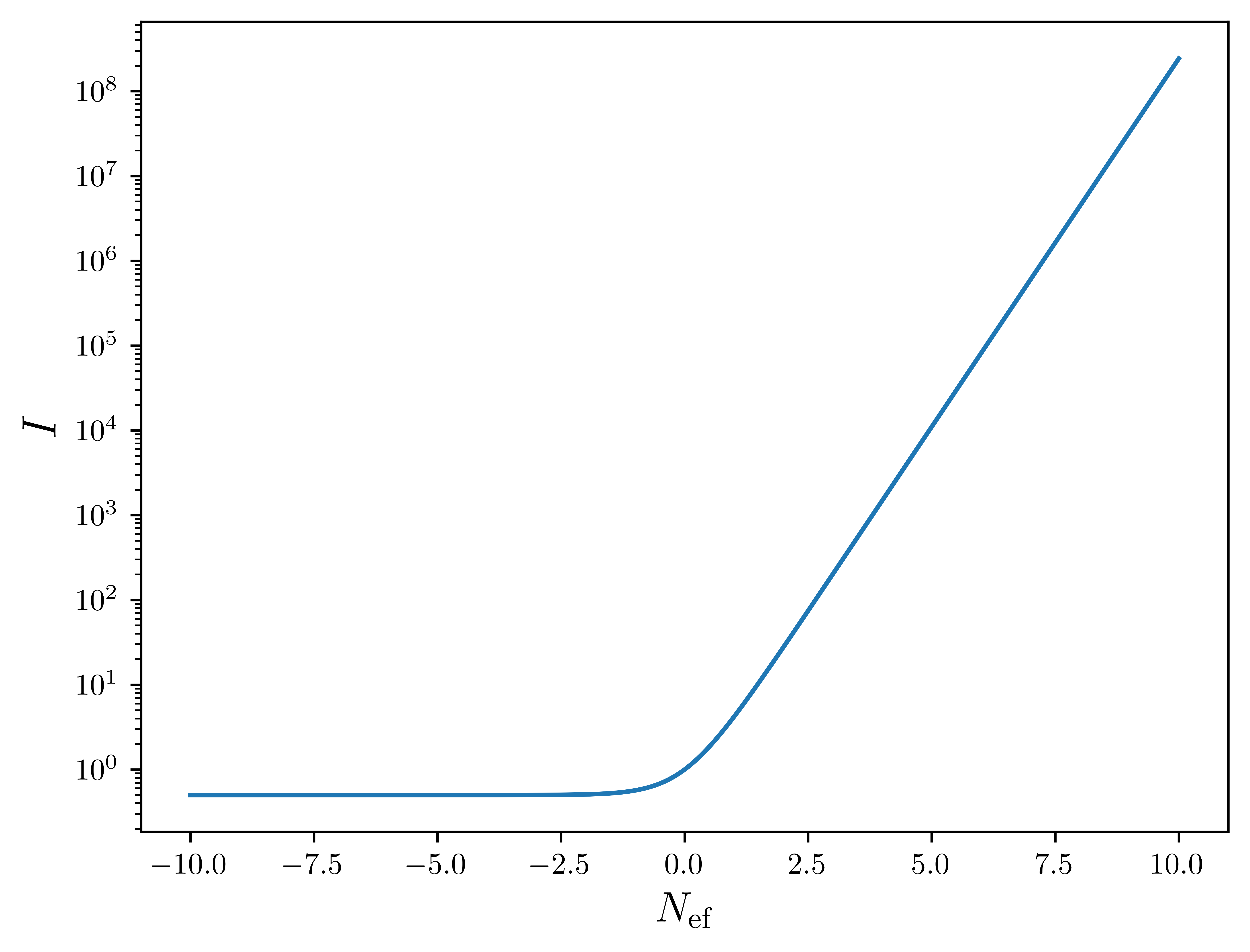}
    \end{minipage}
    \hfill
    \begin{minipage}{0.48\textwidth}
        \centering
        \includegraphics[width=\textwidth]{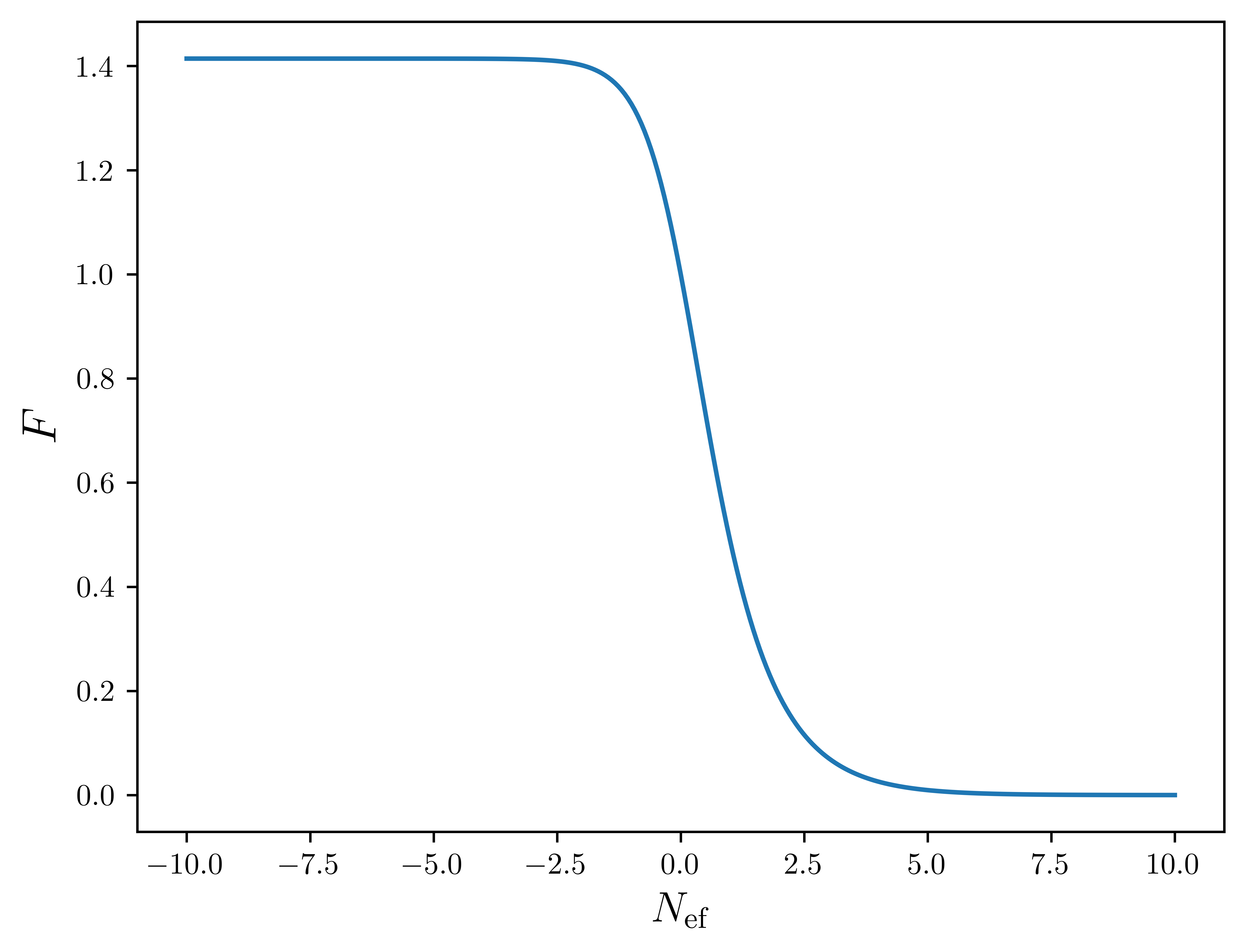}
    \end{minipage}
    \caption{\textit{Left:} Fisher information (in units of $1/\epsilon^2_1$), as a function of the number of \textit{e}-folds. In the far past, the Fisher information is approximately constant and at positive \textit{e}-folds it begins to grow exponentially. \textit{Right:} the function $F(\eta)$, which determines the Cramér-Rao bound. One can notice the variance is lower bounded by a constant value at negative \textit{e}-folds, and unbounded for positive.}
    \label{fig:QFI}
\end{figure}

Using the tools introduced in Sec.~\ref{sec:QFI} one finds the QFI expressed in Eq.~\eqref{eq:QFIfin}.
Substituting the squeezing parameter and the angle found in \eqref{eq:sqref1} and \eqref{eq:sqref2} yields the simple expression
\begin{equation}\label{eq:I}
    I=\frac{1}{2\epsilon_1^2}\left(1+\frac{1}{k^2\eta^2}\right).
\end{equation}
It is remarkable that, despite the complicated expressions found in \eqref{eq:sqref1} and \eqref{eq:sqref2}, Eq.~\eqref{eq:I} simplifies and does not depend on $H$ nor $\Mpl$. This simplification can be made more manifest from the power spectra featured in $\sigma_{\mathcal{R}}$ in Eq.~\eqref{sigmar}. In the super-Hubble regime, i.e. for $|k\eta|\ll 1$, the Fisher information blows up, while in the sub-Hubble regime, $|k\eta|\gg1$, it approaches a constant value of $1/2\epsilon_1^2$. For convenience, we plot in the \textit{Left} panel of Fig. \ref{fig:QFI} the quantum Fisher information $I$ in terms of the number of \textit{e}-folds $N_{ef}=-\log(- k\eta)$.

The Cramér-Rao bound states that
\begin{equation}\label{eq:finCR}
    \Delta\epsilon_1\geq 1/\sqrt{I(\epsilon_1)}\equiv \epsilon_1 F(\eta), \qquad F(\eta) \equiv \sqrt{\frac{2}{1+\frac{1}{k^2\eta^2}}} \in [\sqrt{2},0].
\end{equation}
Notice that on the RHS a true value of $\epsilon_1$ appears. Thus to know the bound on the variance of the parameter, one must first obtain the value with absolute precision. The current conservative constraint on the tensor-to-scalar ratio being of the order $r < 0.036$ \cite{BICEPKeck:2022mhb, Tristram:2020wbi, BICEPKeck:2024stm}, we cannot infer numerical values for the variance.\footnote{At best, by setting $\epsilon_1$ to its upper bound, we find the 
\begin{equation}
    \Delta\epsilon_1\geq 2.25 \times 10^{-3} F(\eta).
\end{equation}} Nevertheless, the time dependence of the QFI is a valuable quantity. Its expression is featured in the \textit{Right} panel of Fig. \ref{fig:QFI}. 
It is notable that in the far past Fisher information is equal to exactly $I = 1/(2\epsilon^2_1)$, leading to the Cramér-Rao bound $\Delta \epsilon \geq \sqrt{2} \epsilon_1$, which remains constant for negative \textit{e}-folds. In this early time regime, the mode of interest is well below the comoving Hubble radius and the squeezing mechanism is mostly ineffective \cite{Grain:2019vnq, Colas:2021llj, Martin:2021znx}. The quantum Fisher information there mostly accounts for the maximal precision one can expect from infering the parameter $\epsilon_1$ in a flat-space vacuum state. Then, around Hubble crossing ($N_{\mathrm{ef}} \sim 0$, i.e. $|k\eta|\sim 1$), $F(\eta)$ transits to zero within a couple of \textit{e}-folds. Physically, the vacuum state is placed into a two-mode squeezed whose statistics seeds the late-time inhomogeneities observed in the CMB and the LSS. It suggests that in this peculiar state, there exists a measurement protocol which can reach exponentially precise inference at late time, $\Delta \epsilon_1 \geq  \sqrt{2} \epsilon_1 \exp (- N_{\mathrm{ef}}) $. Somehow, the squeezing mechanism generates the emergence of a projection that improves the inference of $\epsilon_1$. The next natural question is whether the outcome of this projection is observationally accessible.


\subsection{Optimal POVM \textit{vs} Homodyne Measurement}\label{subsec:optivshomo}

In this Section, we investigate if the observed statistics in the sky is close to the optimal measurement probed by the QFI. 
To achieve this task, we first investigate in Sec.~\ref{subsec:optimal} the optimal POVM, i.e. the one that saturates the  Cramér-Rao bound. We then make a detour to show in Sec.~\ref{subsubsec:optgauss} that the optimal Gaussian measurements are homodyne. At last, we demonstrate in Sec.~\ref{subsec:homodyne} that these homodyne measurements are exponentially suboptimal compare to the optimal POVM found. We conclude that no measurement of a single quadrature is sufficient to reach optimality.

\subsubsection{Optimal POVM}\label{subsec:optimal}

In general, the optimal POVM, i.e. the one that saturates the  Cramér-Rao bound, is given by projectors on the eigenvectors of the symmetric logarithmic derivative \cite{Paris_2003,serafini2024}. Following the notations of Sec.~\ref{sec:QFI}, symmetric logarithmic derivative for a pure state is given by
\begin{equation}\label{eq:SLDdef}
    L_{\vartheta}=2(\ket{\psi_{,\vartheta}}\bra{\psi}+\ket{\psi}\bra{\psi_{,\vartheta}}).
\end{equation}
This operator has rank $2$. Let us express it in an orthonormal basis. $\ket{\psi}$ is already normalized so we can keep it as one of the vectors. The other vector can be constructed as
\begin{equation}
    \ket{\psi^\perp}=\frac{1}{||\psi^\perp||}(\ket{\psi_{,\vartheta}}-\braket{\psi|\psi_{,\vartheta}}\ket{\psi}),
\end{equation}
where
\begin{equation}
    ||\psi^\perp||^2=||\psi_{,\vartheta}||^2-|\braket{\psi|\psi_{,\vartheta}}|^2.
\end{equation}
One can then reexpress Eq.~\eqref{eq:SLDdef} by
\begin{equation}
    L_{\vartheta}=\begin{pmatrix}
        4\Re(\braket{\psi|\psi_{,\vartheta}}) & 2||\psi^\perp||\\
        2||\psi^\perp|| & 0
    \end{pmatrix}.
\end{equation}
We have already computed the relevant entries in \eqref{eq:proof1} and \eqref{eq:proof2}. Bearing in mind that $\bra{\psi_{,\vartheta}}{\psi}\rangle$ is purely imaginary, we reach
\begin{equation}
    L_{\vartheta}=\begin{pmatrix}
        0 & \sqrt{I(\epsilon_1)}\\
        \sqrt{I(\epsilon_1)} & 0
    \end{pmatrix}.
\end{equation}
The eigenvectors are given by 
\begin{equation}\label{sldproj}
    \ket{\pm I(\epsilon_1)}=\frac{1}{\sqrt{2}}(\ket{\psi}\pm \ket{\psi^\perp})=\frac{1}{\sqrt{2}}\left[\left(1\mp\braket{\psi|\psi_{,\vartheta}}\right)\ket{\psi}\pm ||\psi^\perp||^{-1}\ket{\psi_{,\vartheta}}\right].
\end{equation}
Any POVM including projectors on these two eigenvectors is optimal and saturates the quantum Cram\'er--Rao bound. In principle, these von Neumann measurement may be completed arbitrarily on the (infinite dimensional) subspace orthogonal to that spanned by $\ket{\psi}$ and $\ket{\psi_{,\vartheta}}$: all such SLD operators $L_{\vartheta}$ are equivalent in this regard. In our case, Gaussian techniques \cite{monras13,serafini2024,chang25} may be applied to obtain a valid expression for the SLD as a quadratic operator in the canonical field operators. This yields
\begin{equation}
    L_{\vartheta} = r_{,\vartheta}\,\left(\hat{\bf r}^{\sf T} .{\bf M}.\hat{\bf r}\right), \quad {\bf M} \equiv R(\theta/2)Z^{-1/2}\left[\sigma_z-\cosh(2r)\tan(\theta)\sigma_x\right] Z^{-1/2}R(-\theta/2) \; ,
\end{equation}
where $\hat{\bf r}=(\hat{\R}_{\bf k},\hat{\P}_{\bf k})^{\sf T}$ is the vector of canonical operators (so that the expression above is a second-order polynomial in the canonical operators). Observe that, because of the diagonal $\sigma_z$ term, the coupling matrix ${\bf M}$ above is never positive (nor is it negative): therefore, it does not correspond to a mere energy measurement of any pair of canonical coordinates (as, not being positive or negative definite, this coupling matrix cannot be put in Williamson form through a symplectic transformation). 
As such, this is not a measurement that is straightforward to identify, as it would correspond to symplectically rotate the quadratures $\hat{\R}_{\bf k}$ and $\hat{\P}_{\bf k}$ to some new quadratures $\hat{x}_{\bf k}$ and $\hat{y}_{\bf k}$, and then to projecting onto the eigenvectors of some linear combination $a_{\bf k} \hat{x}_{\bf k}^2- b_{\bf k} \hat{y}_{\bf k}^2$ for positive $a_{\bf k}$ and $b_{\bf k}$. In turn, these would correspond to the eigenvectors of an inverted well potential, which may be defined \cite{liuliang} but do not form a resolution of the identity. It is important to remark that, nevertheless, a valid and bounded POVM may be obtained as the projectors on the eigenvectors (\ref{sldproj}), and thus an optimal measurement scheme does in principle exist. Even though such an optimal scheme is difficult to identify in practice, and may not be feasible with standard detection methods, let us emphasise that our main result is the establishment of an absolute lower bound on the variance of the scalar to tensor ratio, which is practically relevant anyway.

Also observe that, as discussed in greater detail in Appendix \ref{2for1}, performing the same measurement on the pair $(\hat{\R}_{-{\bf k}},\hat{\P}_{-{\bf k}})$ would allow one to double the QFI reported above.

\subsubsection{Optimal Gaussian measurement}\label{subsubsec:optgauss}

While the result of the previous section provides an analytical solution for the optimal measurement, it is not exactly satisfying as there is no practical experimental scheme to achieve it. Let us then consider an achievable measurement scheme where we are restricted only to Gaussian measurements, i.e. projections onto Gaussian states of the form $\hat{D}_{\mathbf{r}}.\hat{S}^\dagger(r)\ket{0}$ where $\hat{D}_{\mathbf{r}}$ is the displacement operator defined below Eq.~\eqref{eq:dispop} and $\hat{S}^\dagger(r)$ the squeezing operator defined in Eq.~\eqref{eq:sqtr}. These states are parametrized by a symplectic transformation $S$. We aim at finding the Gaussian measurements optimising the inference.

Given the restriction to a subset of projectors, the optimal inference is given by nothing but the classical Fisher information we now explain how to compute. The probability distribution of $\mathbf{r}$ for a zero-mean single mode state is given by
\begin{equation}
    p(\mathbf{r})=\frac{\exp\left[-\frac{1}{2}\mathbf{r}^T(\sigma+SS^T)^{-1}\mathbf{r}\right]}{2\pi\sqrt{\det{\sigma+SS^T}}}.
\end{equation}
For such a distribution, the classical Fisher information is given by
\begin{equation}
    I_{\mathrm{cl}}=\frac{1}{4}\int \dd^2 \mathbf{r} p(\mathbf{r})\left\{\mathbf{r}^T(\sigma+SS^T)^{-1}\sigma_{,\vartheta}(\sigma+SS^T)^{-1}\mathbf{r}-\Tr[(\sigma+SS^T)^{-1}\sigma_{,\vartheta}]\right\}^2,
\end{equation}
which boils down to evaluating various moments of a Gaussian probability distribution with effective covariance matrix $\sigma+SS^T$. Using Isserlis's theorem (Wick's theorem for probability distributions) one finds
\begin{equation}
    I_{\mathrm{cl}}=\frac{1}{2}\Tr\left\{\left[(\sigma+SS^T)^{-1}\sigma_{,\vartheta}\right]^2\right\}.
\end{equation}
Our goal here is to find the optimal measurement among the set of Gaussian ones, that is we are looking for a symplectic matrix $S$ which extremizes $I_{\mathrm{cl}}$. To achieve this task, we vary $I_{\mathrm{cl}}$ with respect to $S$,
\begin{equation}
    \delta I_{\mathrm{cl}}=2\Tr[S^T(\sigma+SS^T)^{-1}\sigma_{,\vartheta}(\sigma+SS^T)^{-1}\sigma_{,\vartheta}(\sigma+SS^T)^{-1}\delta(S)],
\end{equation}
such that the variational derivative is given by
\begin{equation}\label{eq:var}
    \frac{\delta I_{\mathrm{cl}}}{\delta S}=2S^T(\sigma+SS^T)^{-1}\sigma_{,\vartheta}(\sigma+SS^T)^{-1}\sigma_{,\vartheta}(\sigma+SS^T)^{-1}.
\end{equation}

At last, we look for a matrix $S$ such that $\delta I_{\mathrm{cl}}/\delta S = 0$. If all of the matrices in Eq.~\eqref{eq:var} are invertible, then the solution does not exist, \eqref{eq:var} being a product of non-vanishing matrices. Let us then take a determinant of both sides, using the fact that $\det S^T = 1$, to reach
\begin{equation}
    2\det[\sigma+SS^T]^{-3}\det[\sigma_{,\vartheta}]^2=0.
\end{equation}
The trivial solution $\sigma+SS^T=0$ does not exist for real squeezing parameters, and there is only one family of non-trivial solutions, i.e. the homodyne measurements. They are defined up to a phase as $S=\lim_{z\to\infty}\text{diag}(z,1/z)$, which amounts to only measuring one of the quadratures, that is the first entry of the covariance matrix (up to a phase which allows one to select the quadrature of interest). In this case,
\begin{equation}
    \det[\sigma+SS^T]=2+\frac{H^2}{2k^3\epsilon z^2}+\frac{2k z^2\epsilon}{H^2\eta^2}+\frac{H^2\eta^2}{2kz^2\epsilon},
\end{equation}
which tends to infinity when $z\to\infty$, so its inverse goes to zero. Computing the whole matrix on the left-hand side of Eq.~\eqref{eq:var} and taking $z\to\infty$ yields a zero matrix, that is $\delta I_{\mathrm{cl}}/\delta S = 0$. Thus, the Gaussian measurement that optimises the Fisher information is homodyne.

\subsubsection{Homodyne measurement}\label{subsec:homodyne}

In the previous section, we found that the optimal Gaussian measurement is homodyne. This is actually very good news for observational cosmology, as it is the only measurement to which one can have access. Scales currently probed in the CMB and LSS spent tens of \textit{e}-folds above the comoving horizon before their reentry during radiation and matter domination era. A consequence of Weinberg adiabatic theorem \cite{Weinberg:2003sw} is that correlators of the conjugate momentum $\langle \hat{\mathcal{P}}^n\rangle$ are practically inaccessible, being related to observational quantities exponentially suppressed by the number of \textit{\textit{e}-folds} $N_{\mathrm{ef}}$, see footnote \ref{foot:nef} for details.
Translated into the jargon of homodyne measurements, we can only measure $\langle \hat{\mathcal{R}}^2\rangle$ correlation and the distribution is given by a probability density:
\begin{equation}
    p(x|\epsilon_1)=\frac{1}{\sqrt{2\pi \sigma_{11}(\epsilon_1})}e^{-\frac{x^2}{2\sigma_{11}(\epsilon_1)}},
\end{equation}
where the covariance matrix is specified in Eq. \eqref{eq:covmat}.

A last question subsists: if we restrict ourselves to homodyne measurements, how close are we from the optimal inference found by the QFI in Sec.~\ref{subsec:optimal}? In the homodyne case, the Fisher information simplifies to
\begin{equation}
    I_{\mathrm{cl}}(\epsilon_1)=\int_{-\infty}^{\infty}\dd x p(x|\epsilon_1)\left[\frac{\dd}{\dd \epsilon_1}\log p(x|\epsilon_1)\right]^2=\frac{{\sigma_{11}}'(\epsilon_1)^2}{2\sigma_{11}(\epsilon_1)^2}=\frac{1}{2\epsilon_1^2}.
\end{equation}
Thus:
\begin{equation}\label{eq:homfin}
    \Delta \epsilon_1 \geq \sqrt{2}\epsilon_1, 
\end{equation}
i.e. the approximate value of the bound for number of \textit{e}-folds smaller than zero. On the one-hand, it means that for negative \textit{e}-folds, the homodyne measurement is optimal. On the other hand, it also implies that at late-time, when inflation ends, the only measurement we are in practice able to record in strongly suboptimal compared to the optimal bound found in Eq. \eqref{eq:finCR}.
Note that measuring another direction in phase-space, for instance the momentum direction for which 
\begin{align}
    \widetilde{p}(x|\epsilon_1)=\frac{1}{\sqrt{2\pi \sigma_{22}(\epsilon_1})}e^{-\frac{x^2}{2\sigma_{22}(\epsilon_1)}},
\end{align}
leads to the same result, $I_{\mathrm{cl}}(\epsilon_1) =1 /(2\epsilon_1^2)$.\footnote{Note that it does not mean progress cannot be made on the inference of this parameter. Indeed, the Fisher information informs us about the improvement one can expect from each independent measurement. In cosmology, this corresponds to the understanding of how much error bars shrink when observing new scales or \textit{modes}.} 

The optimal Gaussian measurement turns out to be strongly suboptimal at late time on super-Hubble scales compared to the optimal POVM. Hence, the optimal measurement must necessarily combine quadratures in a way that cannot reduce to the statistics of the Gaussian state along one phase-space axis. Said it differently, one must perform non-Gaussian measurements to infer with precision the parameter $\epsilon_1$. This may not come as a surprise: to probe how squeezed is the state, one needs to assess simultaneously two quadratures \cite{Grain:2019vnq, Martin:2022kph}. Given the smallness of the decaying mode at the end of inflation, this task is known to be exponentially difficult. If one achieves this exponentially difficult measurement, it seems reasonable that the reward is an exponentially enhanced inference. Leaving the feasibility of such experiment aside, the tools developed here provide quantitative arguments to evaluate the gain one may expect from measuring correlators of the conjugate momentum. 

It should also be borne in mind that all of these considerations apply to the pursuit of optimal measurement rates. In practice, one may attain arbitrary precision (at necessarily suboptimal rates) by performing Gaussian state tomography, which can be performed through homodyne measurements alone (but does still require one to rotate the measurement basis, which is already a daunting task in the cosmological context).


\begin{tcolorbox}[%
			enhanced, 
			breakable,
			skin first=enhanced,
			skin middle=enhanced,
			skin last=enhanced,
			before upper={\parindent15pt},
			]{}

            \vspace{0.05in}

           \paragraph{Comparison with observational forecasts.}

Future CMB experiments aim to measure or tightly constrain the tensor-to-scalar ratio $r$ at the pivot scale $k_* = 0.05\,\mathrm{Mpc}^{-1}$ through measurements of the large-scale $B$-mode polarization of the CMB.  
In the Gaussian (Fisher-matrix) approximation, the $1\sigma$ uncertainty on \(r\) is estimated as \cite{Errard_2011, 2011JCAP08001F}
\begin{equation}
\sigma(r) \simeq 
\left[
\sum_{\ell=\ell_{\min}}^{\ell_{\max}}
\frac{(2\ell+1)f_{\rm sky}}{2}
\frac{\big(\partial C_\ell^{BB}/\partial r\big)^2}
{(C_\ell^{BB}+N_\ell)^2}
\right]^{-1/2}.
\end{equation}
Here, \(\ell\) denotes the CMB multipoles.
The limits \(\ell_{\min}\) and \(\ell_{\max}\) specify the range of multipoles included in the analysis, typically from \(\ell_{\min}\!\sim\!2\) to \(\ell_{\max}\!\sim\!200\) for large-scale \(B\)-mode forecasts.
The factor \(f_{\rm sky}\) is the observed sky fraction (\(0<f_{\rm sky}\le1\) typically \(f_{\rm sky}\sim0.5\)--\(0.7\) for future CMB experiments).
The quantity \(C_\ell^{BB}\) is the theoretical \(B\)-mode angular power spectrum, which includes both the primordial tensor contribution and the lensing component, while \(N_\ell\) represents the effective noise power spectrum, including instrumental noise, residual foregrounds, and any remaining lensing \(B\)-modes after delensing.
The derivative \(\partial C_\ell^{BB}/\partial r\) describes how the \(B\)-mode power spectrum varies with the tensor-to-scalar ratio \(r\). 

Representative projected \(1\sigma\) sensitivities from current forecast studies are as follows.
The Simons Observatory (SO) aims for \(\sigma(r)\simeq2\times10^{-3}\)~\cite{SimonsObservatory:2018koc}.
The LiteBIRD satellite targets a precision of \(\sigma(r)\simeq(6\text{--}10)\times10^{-4}\),
depending on delensing efficiency and foreground assumptions~\cite{LiteBIRD:2020khw, LiteBIRD:2022cnt}.
CMB-S4 is expected to reach \(\sigma(r)\simeq5\times10^{-4}\)~\cite{CMB-S4:2020lpa},
while the proposed PICO satellite could achieve \(\sigma(r)\lesssim10^{-4}\)
under optimistic assumptions~\cite{NASAPICO:2019thw}.
If a positive detection is made, the expected fractional precision is thus at the level of
a few tens of percent for \(r\sim10^{-3}\) with LiteBIRD or CMB-S4,
and down to a few percent for \(r\gtrsim5\times10^{-3}\).

Let us compare these results with an estimate of the optimal inference from QFI, $  \sigma_{\mathrm{opt}} (r) $. The pivot scale $k_* = 0.05\,\mathrm{Mpc}^{-1}$ typically spends around $N_{\mathrm{ef}} \sim 50-55$ \textit{e}-folds above the horizon during typical inflationary models. Assuming a detection at $r \sim 10^{-3}$, we can deduce $\sigma_{\mathrm{opt}} (r) $ from $\Delta \epsilon_1 \geq  \sqrt{2} \epsilon_1 \exp (- N_{\mathrm{ef}}) $ and the relation $r = 16 \epsilon_1$. We obtain
\begin{align}
    \sigma_{\mathrm{opt}}  (r) \gtrsim r . \exp (- N_{\mathrm{ef}}) \sim 10^{-25}.
\end{align}
This estimate only captures the contribution of one single independent measurement. In order to obtain a realistic estimate, one further needs to divide by the number of independent measurements (see Eq. \eqref{eq:estim}), 
\begin{align}
    N = \sum_{\ell = 2}^{200} (2 \ell + 1)^2 \sim 10^{7},
\end{align}
leaving $\sigma^{\mathrm{CMB}}_{\mathrm{opt}}  (r) = \sigma_{\mathrm{opt}}  (r) / N \sim  10^{-32}$.
Compared to the optimistic estimates of $\sigma(r)\simeq10^{-4}$, there is still room for about $28$ orders of magnitude before saturating to optimal inference! 

At last, if one considers instead the optimal homodyne measurement explored in Sec. \ref{subsec:optivshomo}, one finds from Eq. \eqref{eq:homfin} the one-mode optimal inference $\sigma_{\mathrm{hom}}  (r) \gtrsim \sqrt{2} r \sim 10^{-3}$, much closer to the bounds from future experiments. Yet, if one further accounts for the number of independent modes $N$, this yields $\sigma^{\mathrm{CMB}}_{\mathrm{hom}}  (r) = \sigma_{\mathrm{hom}}  (r) / N \sim  10^{-10}$, leaving at least $6$ orders of magnitude improvements for future experiments before saturating the bound.

\end{tcolorbox}

\section{Conclusion}\label{sec:conclu}


Understanding the extent to which we can reconstruct the mechanism behind the generation of cosmological inhomogeneities in the early universe is crucial for assessing the inflationary paradigm.
%
%
In this work, we applied a quantum metrology method — the quantum Fisher information — to the squeezed state describing cosmological inhomogeneities at the end of inflation. The QFI quantifies the optimal precision achievable in inferring a microphysical parameter if arbitrary measurements on inflationary fluctuations were possible. To this end, we derived an expression for the QFI in terms of the squeezing parameters of the state, as given in Eq.~(\ref{eq:QFIfin}), which directly relates derivatives of the squeezing parameter with respect to the parameter of interest to the QFI. 

Applying this formalism to the inflationary squeezed states, we focused on the inference of the tensor-to-scalar ratio $r$, which is directly related to the first slow-roll parameter $\epsilon_1$. Our main result, shown in Fig.~\ref{fig:QFI}, reveals an exponential growth of the QFI, $I \propto \exp(2 N_{\mathrm{ef}})$, on super-Hubble scales. This implies that, in principle, access to all possible measurements would allow exponentially precise inference of $r$.

However, cosmological observations are limited to measuring only one entry of the covariance matrix $\sigma_{\mathcal{R}}$ — namely, the power spectrum of curvature perturbations. The practical impossibility of measuring the power spectra associated with the conjugate momentum restricts us to homodyne measurements along a single quadrature direction. We computed the classical Fisher information under these measurement constraints and compared it to the QFI. Unlike the QFI, the classical Fisher information remains constant throughout inflation, equal to the initial value of the QFI. This holds not only for homodyne measurements along the position quadrature (accessible in cosmology) but also for any measurement along a single phase-space direction.

    These results indicate that although late-time squeezing can greatly enhance inference precision in principle, extracting this information requires non-Gaussian measurements that simultaneously access both quadratures, which is beyond the reach of current cosmological experiments. For that purpose, one would need to access an operator that involves the so-called decaying mode which is notoriously hard. Moreover, this observation has to be done through a highly non-trivial POVM. \\


This study highlights once again the peculiar nature of the quantum state describing cosmological perturbations. On one hand, these highly squeezed states enable extremely precise parameter inference thanks to the strong squeezing of their quadratures. On the other hand, the practical limitation of experiments to homodyne measurements precludes any such improvement in real data analysis. This tension illustrates another facet of the ambiguity surrounding squeezing and the quantum-to-classical transition~\cite{Grain:2019vnq, Colas:2021llj, Colas:2023wxa}: while the quantum states of cosmological inhomogeneities exhibit pronounced non-classical features, the restricted observational access during inflation confines us to probing an effectively classical probability distribution, devoid of unmistakable quantum signatures. 

The idea that part of the information encoded in cosmological perturbations becomes inaccessible due to the decaying mode has been discussed since the early developments of the squeezing formalism \cite{Grishchuk:1990bj,Grishchuk:1992tw,Albrecht:1992kf,Polarski:1995jg}. What has been missing, however, is a quantitative characterization of this information loss in terms of well-defined quantum information measures. The Quantum Fisher Information provides a general framework based on metrology to address this question and to quantify, in a precise and operational way, how much information is lost. It is reassuring that this quantitative analysis confirms the qualitative picture suggested by earlier works — even if it renders the result less striking. At the same time, our construction establishes a general formalism that can be extended to other physically relevant scenarios (e.g. ultra-slow roll), where the outcome of this computation may reveal more distinctive or unexpected behavior.

It remains frustrating that many approaches — including the one pursued in this work — to uncover the quantum origin of cosmological inhomogeneities ultimately rely on accessing the decaying mode, which is effectively unobservable. While one possibility is to continue hoping that such a measurement may one day become feasible, this tends to push us toward finely tuned scenarios, such as very small scales or departures from slow roll. Moreover, much of the existing literature, ours included, relies heavily on Gaussian statistics. Developing strategies that do not hinge on detecting the decaying mode therefore appears essential. In this sense, exploiting the shape dependence of the primordial bispectrum is a promising direction, and we hope future work will continue to build on this idea \cite{Green:2020whw, Salcedo:2025ezu}.

Beyond observational considerations, we also expect quantum information methods to shed light on theoretical aspects of primordial cosmology. Techniques aimed at constraining EFT parameters using quantum information measures — for instance purity or entanglement entropy \cite{DuasoPueyo:2023viy, Cai:2025kcd} — or using quantum speed limits may offer new avenues for addressing the challenges posed by Lorentz-breaking theories. Even if the decaying mode remains practically inaccessible, the phase of the wavefunction retains information that could be relevant for such conceptual investigations. A more systematic understanding of how different quantum-estimation techniques relate and differ — for example, the connection between complexity and purity \cite{Zhou:2025ahz} — is an important step toward a cohesive formulation of quantum field theory in the early universe.\\


Classical Fisher information is widely used in cosmological forecasts for upcoming experiments. Extending the Quantum Fisher Information results presented here to forecast analyses would help quantify the information lost by restricting measurements to the curvature-curvature power spectrum. Applying these methods to infer other microphysical parameters, such as the Hubble parameter $H$, could provide further insights. Finally, investigating whether the squeezing parameters themselves can be directly inferred — rather than the underlying microphysical parameters — would be valuable, as these parameters carry rich information about the inflationary era.

Throughout this work we have restricted ourselves to a simple inflationary model; however, the formalism presented in this paper remains applicable and can be extended to a much broader class of theories, as long as one works within linear theory, with the covariance matrix appropriately adjusted. Some natural extension of it is the  multifield scenarios \cite{Langlois:2008mn, Chen:2009zp, Cespedes:2012hu, Arkani-Hamed:2015bza, Dias_2018, Masoumi_2017}, which stems from intersection of string theory and inflationary cosmology. The model describes a multiverse with a population of vacua given by an energy landscape which determines the possibility of quantum tunnelling between vacuum states. Then, one can consider the introduction of slow-roll corrections \cite{Boyanovsky_2006, Martin:2024ofv}. A particularly interesting scenario departing from these corrections corresponds to the ultra slow-roll case~\cite{Dimopoulos:2017ged,Pattison:2018bct, Vennin:2020kng}. Exploring these scenarios can require the release of the Gaussianity assumption and explore Non-Gaussianities~\cite{Chen:2010xka} in a perturbative way in which our method is still valid.

Beyond these ideas, a particularly interesting direction concerns back-reaction effects, which can be investigated within the framework of stochastic inflation by splitting the inflaton field into a classical stochastic component — obtained by coarse-graining the quantum field — and a remaining, manifestly quantum contribution~\cite{Levasseur_2015}. The stochastic part enters as Gaussian noise adding on a total covariance matrix, making it naturally compatible with Gaussian quantum information approaches, such as diffusive quantum mechanics (see \cite{serafini2024} for an application in Gaussian quantum information).

Another natural extension would be to investigate how our results are modified in scenarios that provide alternatives to inflation. In particular, \cite{Colas:2024xjy} studies a model that interpolates between inflation and ekpyrosis — a bouncing–universe scenario for the early universe. The distinct physical mechanisms responsible for generating cosmological inhomogeneities yield different expectations for the inferred tensor-to-scalar ratio, and these differences are cleanly captured by the QFI formalism developed in this paper.

Let us conclude by highlighting that the framework developed throughout this paper can be applicable to many other problems in field theory as long as the considered transformations are symplectic and the states are Gaussian. For example, in the Unruh effect, the modes of frequency $\omega$ are accelerated with a squeezing transformation controlled by $\tanh r=e^{-\pi\omega c/A}$ with $A$ being the acceleration related to the temperature \cite{Hu:2018psq, Aspachs:2010hh}. Similar transformations also appear in black hole physics \cite{PhysRevD.34.373,Agullo2023, Calamanciuc2023, Bradler2014, Wu2022}, as well as in more experimentally feasible phenomena like Schwinger particle production \cite{Li_2017}. We leave these exciting perspectives for future works. 

\subsection*{Acknowledgements:} We thank William R. Coulton, Daniel Green, J\'er\^ome Martin, Amaury Micheli, Oliver Philcox, and Vincent Vennin for their useful comments. AA-S is funded by the Deutsche Forschungsgemeinschaft (DFG, German Research Foundation) — Project ID 516730869. A.S. thanks M.G.~Genoni and M.G.A. Paris for recurrent discussions and consultancies on quantum estimation theory. T.C. thanks the  organizers and participants of the 2025 Peyresq Spacetime Meeting, in particular Patricia Ribes Metidieri and Don Page, together with the Julian Schwinger Foundation for financial support of the workshop. This work has been supported by STFC consolidated grant ST/X001113/1, ST/T000694/1, ST/X000664/1 and EP/V048422/1. This work was supported by the Engineering and Physical Sciences Research Council [grant number EP/S021582/1] and partially supported through Grant  No.  PID2023-149018NB-C44 (funded by MCIN/AEI/10.13039/ 501100011033). AS also acknowledges funding from the Leverhulme Trust Research Project Grant RPG-2024-28. 


\appendix

\section{Scalar field quantization}
\label{app:app0}

In this Appendix, we construct the Fock space of a single scalar field in flat spacetime to illustrate the quantisation procedure adopted in the main text for the cosmological scalar field. Let us consider 
\begin{equation}
    S=\int\dd^4x\mathcal{L}=\frac{1}{2}\int \dd^4x \left[(\nabla\phi)^2-m^2\phi^2\right]=\frac{1}{2}\int \dd^4x\left[(\partial_t\phi)^2-\partial_{i}\phi\partial^i\phi - m^2\phi^2\right].
\end{equation}
Extremising this action yields Euler-Lagrange equations in the form of the Klein-Gordon equation
\begin{equation}
    \Box\phi+m^2\phi=0.
\end{equation}
The solution is in terms of plain waves $e^{\pm p_\mu x^\mu}$ where $p^\mu$ is a four-momentum and $x^\mu$ spans time and space. It is easy to identify positive and negative norm modes and express the general solution as
\begin{equation}
    \phi = \int \frac{\dd^3 \boldsymbol{p}}{(2\pi)^3} \left[ a(\boldsymbol{p}) e^{-i p_\mu x^\mu} + a^\dagger(\boldsymbol{p}) e^{+ip_\mu x^\mu} \right],
\end{equation}
with a dispersion relation $p_\mu p^\mu =- m^2$. Upon quantization, the Fourier coefficients become creation/annihilation operators and are used to build the Fock space through
\begin{equation}
    \mathcal{F}=\bigoplus_m \mathcal{F}_m,
\end{equation}
where $\mathcal{F}_m$ corresponds to the $m-$particle space of symmetric (bosonic) square-integrable functions \cite{Schwarz:2020dwt}. The basis states on the Fock space are labelled by $\ket{m}$ and satisfy
\begin{align}
\begin{split}
    &\hat{a}({\boldsymbol{p}})\ket{0}=0,\\
    &\hat{a}({\boldsymbol{p}})\ket{m}=\sqrt{m}\ket{m-1},\\
    &\hat{a}^\dagger({\boldsymbol{p}})\ket{m}=\sqrt{m+1}\ket{m+1}.
\end{split}
\end{align}
These operators also inherit an algebraic structure from the Poisson bracket relations of the form
\begin{equation}
    [a(\boldsymbol{p}),a^\dagger(\boldsymbol{p})]=\mathds{1}.
\end{equation}
In this paper, we shall quantize the relevant fields in an equivalent, although slightly different, manner. Let us go back to the original action and consider canonical momenta conjugate to the field $\phi$. It is given by
\begin{equation}
    \pi=\frac{\partial\mathcal{L}}{\partial\partial_t\phi}=\partial_t\phi.
\end{equation}
Using the Legendre transformation, we can find the Hamiltonian
\begin{equation}
    H=\int\dd^3x\mathcal{H}=\int\dd^3x(\pi\partial_t\phi-\mathcal{L})=\frac{1}{2}\int\dd^3x(\pi^2+\partial_i\phi\partial^i\phi+m^2\phi).
\end{equation}
Let us perform a symplectomorphism in the form of a Fourier transform
\begin{align}
\begin{split}
    &\phi(t,x)=\int_{\mathds{R}^{3+}}\frac{\dd^3 \boldsymbol{k}}{(2\pi)^3} \phi_{\boldsymbol{k}}(t)e^{-i\boldsymbol{k}.\boldsymbol{x}},\\
    &\pi(t,x)=\int_{\mathds{R}^{3+}}\frac{\dd^3 \boldsymbol{k}}{(2\pi)^3} \pi_{\boldsymbol{k}}(t)e^{-i\boldsymbol{k}.\boldsymbol{x}}.
\end{split}
\end{align}
The Hamiltonian is thus
\begin{equation}
    H=\int_{\mathds{R}^{3+}}\frac{\dd^3 \boldsymbol{k}}{(2\pi)^3}\left[\pi_{\boldsymbol{k}}\pi_{-\boldsymbol{k}}+\left(k^2+m^2\right)\phi_{\boldsymbol{k}}\phi_{-\boldsymbol{k}}\right].
\end{equation}
The symplectic form in new coordinates is given by
\begin{equation}
	\omega=\int_{\mathds{R}^{3+}}\frac{\dd^3 \boldsymbol{k}}{(2\pi)^3} \mathbf{d}\phi_{\boldsymbol{k}}\wedge\mathbf{d}\pi_{-\boldsymbol{k}}.
\end{equation}
We can find the equations of motion using Hamiltonian flow,
\begin{align}
\begin{split}
\label{eq:scalar_eom}
    &\dot{\phi}_{\boldsymbol{k}}=\omega(X_{\phi_{\boldsymbol{k}}},X_H)=\pi_{\boldsymbol{k}},\\
    &\dot{\pi}_{\boldsymbol{k}}=\omega(X_{\phi_{\boldsymbol{k}}},X_H)=-(k^2+m^2)\phi_{\boldsymbol{k}},
\end{split}
\end{align}
which is naturally equivalent to the Klein-Gordon equation. The solution can be written as
\begin{equation}
\label{eq:symplev}
    \begin{pmatrix}
        \phi_{\boldsymbol{k}}(t)\\
        \pi_{\boldsymbol{k}}(t)
    \end{pmatrix}=\begin{pmatrix}
        \alpha_{\boldsymbol{k}}(t) & \beta_{\boldsymbol{k}}(t)\\
        \beta_{\boldsymbol{k}}^*(t) & \alpha_{\boldsymbol{k}}^*(t)
    \end{pmatrix}\begin{pmatrix}
        \phi_{\boldsymbol{k}}(0)\\
        \pi_{\boldsymbol{k}}(0)
    \end{pmatrix},
\end{equation}
where $|\alpha_{\boldsymbol{k}}(t)|^2-|\beta_{\boldsymbol{k}}(t)|^2=1$. The elements $\alpha_{\boldsymbol{k}}(t),\beta_{\boldsymbol{k}}(t)$ are called the Boguliubov coefficients and they satisfy the same differential equation as $\phi_{\boldsymbol{k}},\pi_{\boldsymbol{k}}$. The matrices of coefficients above form a complex representation of the real symplectic group $Sp(2,\mathds{R})$ (to which they are isomorphic through a unitary change of basis). Indeed, one may recover the usual Fock space by considering a symplectic transformation of the form
\begin{align}
\begin{split}
    &\phi_{\boldsymbol{k}}=\frac{1}{\sqrt{2 k}}\left[a_{\boldsymbol{k}}(t)+a^\dagger_{-\boldsymbol{k}}(t)\right], \qquad
    \pi_{\boldsymbol{k}}=-i\sqrt{\frac{k}{2}}\left[a_{\boldsymbol{k}}(t)-a^\dagger_{-\boldsymbol{k}}(t)\right],
\end{split}
\end{align}
and quantizing $a_{\boldsymbol{k}},\, a^\dagger_{\boldsymbol{k}}$. Of course, one may move between the solution in terms of creation/annihilation coefficients and the quadrature operators freely using the above-defined transformation, and the symplectic evolution will differ by matrix congruence. 

\section{Embedded single-mode structure}\label{2for1}

The single-mode treatment we employ throughout the paper applies even though the states we are dealing with are actually two-mode entangled states, between the ${\boldsymbol{k}}$ and $-{\boldsymbol{k}}$ modes. This follows from an embedding of a single-mode symplectic algebra that is straightforward but not quite trivial, and thus worthwhile elucidating here.

As mentioned in the main text, the operator $\hat{v}_{\boldsymbol{k}}$ is actually canonically conjugated to $\hat{p}_{-\boldsymbol{k}}$; thus, under the ordering ($\hat{v}_{\boldsymbol{k}}$,$\hat{p}_{\boldsymbol{k}}$,$\hat{v}_{-\boldsymbol{k}}$,$\hat{p}_{-\boldsymbol{k}}$), the two-mode symplectic form reads
\begin{equation}\label{omega2}
\Omega = \left(\begin{array}{cc}
0 & \omega \\
\omega & 0
\end{array}
\right) 
\end{equation}
(where $0$ stands for the $2\times2$ null matrix and $\omega$ for the $2\times2$ anti-symmetric symplectic form). It is now clear that, in this canonical basis, linear transformations of the form $S\oplus S$, where $S$ is a $2\times2$ one-mode symplectic, preserve the two-mode symplectic form $\Omega$ and are therefore symplectic. These are transformations that act in the same way on the pairs $\{\hat{v}_{\boldsymbol{k}},\hat{p}_{\boldsymbol{k}}\}$ and $\{\hat{v}_{-\boldsymbol{k}},\hat{p}_{-\boldsymbol{k}}\}$. Indeed, transformations like those of Eq.~(\ref{symp}), which describe the inflationary dynamics, do not depend on the sign of $\boldsymbol{k}$. By the same token, Eq.~(\ref{class}) describes an actual two-mode symplectic transformation, since $\hat{v}_{-\boldsymbol{k}}$ and $\hat{p}_{-\boldsymbol{k}}$ transform the same way under them. 

Note also that, by applying the singular value decomposition to symplectic transformations of the form $S\oplus S$, one can see that our two-mode pure states are, in standard quantum optical terminology, single-mode states squeezed in orthogonal phase-space directions undergoing a beam-splitting transformation.

\paragraph{Consequences for quantum estimation.}

This embedded single-mode structure can be maintained not only in describing the dynamics, but also for the evaluation of the QFI and, with a little care, of the symmetric logarithmic derivative operator that determines the optimal projective measurement. 

As already remarked in the main text, for the QFI evaluation, the characteristic function approach of Eq.~(\ref{chara}) shows that, for two-mode states with CM of the form $\sigma\oplus \sigma$, as are ours, the QFI is clearly additive (notice that this is the case even if the overall state is not a tensor product of two single-mode states). The value we presented must hence be doubled if access to all four variables is taken into account.

As for the symmetric logarithmic derivative, it should be mentioned that, for pure states, its Gaussian formula hinges on the determination of eigenmatrices of the linear map $A \mapsto \Omega A \Omega, \: \forall A$ (left and right multiplication by the symplectic form $\Omega$) \cite{serafini2024}. For our pure states with CM of the form $\sigma\oplus \sigma$, it can be seen that the only two relevant eigenmatrices are, in terms of standard Pauli matrices, $\sigma_x\oplus\sigma_x$ and $\sigma_z\oplus\sigma_z$. This effectively reduces the problem of determining the SLD operator to a single-mode one, with the only caveat that the full (doubled) QFI can be harnessed by performing the same optimal measurement on the $-{\bf k}$ modes too. 


\section{Quantum Fisher information from states in the Fock space}\label{app:app2}

In this Appendix, we provide an alternative derivation of the QFI in terms of the squeezing parameters \eqref{eq:QFIfin} using the Fock's space representation of the state. Recall the form of the symplectic transformation
\begin{equation}
    S(r,\theta)=R(\theta/2)S(r)R(-\theta/2)=R(\theta/2)\begin{pmatrix}
        e^r & 0\\
        0 & e^{-r}
    \end{pmatrix}R(-\theta/2).
\end{equation}
The vacuum is rotationally invariant, thus
\begin{equation}
    \hat{R}(-\theta/2)\ket{0}=\ket{0}.
\end{equation}
Squeezing acts on it as
\begin{equation}
    \hat{S}(r)\ket{0}=\frac{1}{\sqrt{\cosh(r)}}\sum_{m=0}^{\infty}\left(\frac{\tanh(r)}{2}\right)^m\frac{\sqrt{2m!}}{m!}\ket{2m}.
\end{equation}
The embedding detailed in Appendix \ref{2for1} reveals that the single-mode squeezing in question is actually acting on the Fock space pertaining to the canonical pair $\hat{R}_{\bf k}$ and $\hat{\P}_{-{\bf k}}$, to which the Fock states $\ket{2m}$ above belong. 
Now let us rotate it to obtain the squeezed state:
\begin{equation}
   \ket{\psi}=e^{-i\theta/2 a^\dagger a} \hat{S}(r)\ket{0}=\frac{1}{\sqrt{\cosh(r)}}\sum_{m=0}^{\infty}\left(\frac{e^{-i\theta}\tanh(r)}{2}\right)^m\frac{\sqrt{2m!}}{m!}\ket{2m}.
\end{equation}

We now need to evaluate how this state varies when we transform a parameter $\vartheta$. The derivative with respect to $\vartheta$ is given by
\begin{align}
\begin{split}
    \ket{\psi_{,\vartheta}}&=\frac{r_{,\epsilon_1}}{-2\cosh^{3/2}(r)}\sinh(r)\sum_{m=0}^{\infty}\left(\frac{e^{-i\theta}\tanh(r)}{2}\right)^m\frac{\sqrt{2m!}}{m!}\ket{2m}\\&+\frac{1}{\sqrt{\cosh(r)}}\sum_{m=0}^{\infty}(-im\theta_{,\epsilon_1})\left(\frac{e^{-i\theta}\tanh(r)}{2}\right)^m\frac{\sqrt{2m!}}{m!}\ket{2m}\\
    &+\frac{1}{\sqrt{\cosh(r)}}\frac{r_{,\epsilon_1}}{\cosh(r)\sinh(r)}\sum_{m=0}^{\infty}m\left(\frac{e^{-i\theta}\tanh(r)}{2}\right)^m\frac{\sqrt{2m!}}{m!}\ket{2m},\\
\end{split}
\end{align}
which simplifies to 
\begin{align}
	\begin{split}
		\ket{\psi_{,\vartheta}}&=-\frac{r_{,\epsilon_1} \tanh(r)}{2}\ket{\psi}+\frac{1}{\sqrt{\cosh(r)}}\sum_{m=0}^{\infty}(-im\theta_{,\epsilon_1})\left(\frac{e^{-i\theta}\tanh(r)}{2}\right)^m\frac{\sqrt{2m!}}{m!}\ket{2m}\\
		&+\frac{1}{\sqrt{\cosh(r)}}\frac{r_{,\epsilon_1}}{\cosh(r)\sinh(r)}\sum_{m=0}^{\infty}m\left(\frac{e^{-i\theta}\tanh(r)}{2}\right)^m\frac{\sqrt{2m!}}{m!}\ket{2m}.\\
	\end{split}
\end{align}

To evaluate the QFI, let us start by computing overlap of the varied state $\ket{\psi_{,\vartheta}}$ with $\ket{\psi}$
\begin{align}
\begin{split}
    \braket{\psi|\psi_{,\vartheta}} &= -\frac{r_{,\epsilon_1} \tanh(r)}{2} - \frac{i\theta_{,\epsilon_1}}{\cosh(r)}\sum_{m=0}^{\infty}m\left(\frac{\tanh(r)}{2}\right)^{2m}\frac{2m!}{(m!)^2}\\&+\frac{r_{,\epsilon_1}}{\cosh^2(r)\sinh(r)}\sum_{m=0}^{\infty}m\left(\frac{\tanh(r)}{2}\right)^{2m}\frac{2m!}{(m!)^2}.
\end{split}
\end{align}
To compute the series, let us use the normalization
\begin{equation}
    \braket{\psi|\psi}=\frac{1}{\cosh(r)}\sum_{m=0}^{\infty}\left(\frac{\tanh(r)}{2}\right)^{2m}\frac{2m!}{(m!)^2}=1,
\end{equation}
such that we have the series
\begin{align}
\begin{split}
    S_1&=\sum_{m=0}^{\infty}m\left(\frac{\tanh(r)}{2}\right)^{2m}\frac{2m!}{(m!)^2}=\sum_{m=1}^{\infty}\left(\frac{\tanh(r)}{2}\right)^{2m}\frac{2m!}{(m-1)!m!}\\&=\sum_{m=0}^{\infty}\left(\frac{\tanh(r)}{2}\right)^{2m+2}\frac{[2(m+1)]!}{(m+1)!m!}=\sum_{m=0}^{\infty}\left(\frac{\tanh(r)}{2}\right)^{2m+2}(4m+2)\frac{2m!}{(m!)^2}.
\end{split}
\end{align}
Reshuffling terms from left and right side and collecting powers of $m$ gives
\begin{equation}
    S_1=\sum_{m=0}^{\infty}m\left(\frac{\tanh(r)}{2}\right)^{2m}\frac{2m!}{(m!)^2}=\frac{\tanh^2(r)\cosh(r)}{2(1-\tanh^2(r))}=\frac{1}{2}\cosh(r)\sinh^2(r).
\end{equation}
Thus,
\begin{equation}
     \braket{\psi|\psi_{,\vartheta}} = -\frac{r_{,\epsilon_1} \tanh(r)}{2} - \frac{i\theta_{,\epsilon_1}}{2}\sinh^2(r)+\frac{r_{,\epsilon_1}\tanh(r)}{2}=- \frac{i\theta_{,\epsilon_1}}{2}\sinh^2(r).
\end{equation}
To wrap up the computation, we also need the sum
\begin{equation}
    S_2=\sum_{m=0}^{\infty}m^2 \left(\frac{\tanh(r)}{2} \right)^{2m}\frac{(2m)!}{(m!)^2}.
\end{equation}
Using the same method,
\begin{align}
\begin{split}
    S_2&=\sum_{m=0}^{\infty}m^2\left(\frac{\tanh(r)}{2}\right)^{2m}\frac{2m!}{(m!)^2}=\sum_{m=0}^{\infty}\left(\frac{\tanh(r)}{2}\right)^{2m+2}(4m^2+6m+2)\frac{2m!}{(m!)^2},
\end{split}
\end{align}
such that
\begin{equation}
    S_2=\sum_{m=0}^{\infty}m^2\left(\frac{\tanh(r)}{2}\right)^{2m}\frac{2m!}{(m!)^2}=\tanh^2(r)S_2+\frac{3}{2}\tanh^2(r)S_1+\frac{1}{2}\tanh^2(r)S_0,
\end{equation}
where $S_0=\sum_{m=0}^{\infty}\left(\frac{\tanh(r)}{2}\right)^{2m}\frac{2m!}{(m!)^2}$. It leads to
\begin{align}
\begin{split}
    S_2&=\frac{\tanh^2(r)}{2(1-\tanh^2(r))}(3S_1+S_0)=\frac{1}{8}\cosh(r)(1+3\cosh(2r))\sinh^2(r).
\end{split}
\end{align}
Note that there exists an alternative way of computing both of these sums. One notes the generating function:
\begin{equation}
    \sum_{m=0}^{\infty} x^m \frac{(2m)!}{(m!)^2} = \frac{1}{\sqrt{1-4x}}, \quad |x|<\frac{1}{4}.
\end{equation}
By taking derivatives with respect to $x$, subsequently multiplying by $x$, and substituting $x=(\tanh(r)/2)^2$, one may contract any sum of this form as long as $|x|<1/4$. The only issues are when $r\to\infty$ as then $x\to1/4$, and more care is required then.\\

We can now compute the norm of the derivative:
\begin{align}
    ||\psi_{,\vartheta}||^2&=\frac{1}{4}r_{,\epsilon_1}^2\tanh^2(r)+\frac{\theta_{,\epsilon_1}^2}{\cosh(x)}S_2+\frac{r_{,\epsilon_1}^2}{\cosh^3(r)\sinh^2(r)}S_2-\frac{r_{,\epsilon_1}^2}{\cosh^{2}(r)\sinh(r)}\tanh(r)S_1 \nonumber \\
    &=\frac{1}{2}r_{,\epsilon_1}^2+\frac{1}{8}(1+3\cosh(2r))\sinh^2(r)\theta_{,\epsilon_1}^2.
\end{align}
And thus, the norm of the normal vector is
\begin{equation}
    ||\psi^\perp||^2=||\psi_{,\vartheta}||^2-|\braket{\psi|\psi_{,\vartheta}}|^2=\frac{1}{2}\left[r_{,\epsilon_1}^2+\cosh^2(r)\sinh^2(r)\theta_{,\epsilon_1}^2\right].
\end{equation}
This yields Fisher information
\begin{equation}
    I=4 ||\psi^\perp||^2 =2\left[r_{,\epsilon_1}^2+\cosh^2(r)\sinh^2(r)\theta_{,\epsilon_1}^2\right].
\end{equation}
We can use the requirement that off-diagonal elements of the derivative of covariance matrix go to zero and find
\begin{equation}\label{eq:rel}
    I=2(r_{,\epsilon_1}^2+\cosh^2(r)\sinh^2(r)\theta_{,\epsilon_1}^2)=2\left[\cosh^2(2r)\tan^2(\theta)+1\right]r_{,\epsilon_1}^2.
\end{equation}
This agrees with our main text result.


%
\bibliographystyle{JHEP}
\bibliography{biblio}

\end{document}